\documentclass{emulateapj}

\slugcomment{Accepted for publication in the Astrophysical Journal }

\shorttitle{Equation of state dependence of nonlinear mode-tide coupling in coalescing binary neutron stars}
\shortauthors{Yixiao Zhou and Fan Zhang}

\usepackage{amsmath,amssymb}
\usepackage[normalem]{ulem}
\usepackage{bm}
\usepackage{comment}
\usepackage{xcolor}
\usepackage{soul}
\usepackage{hyperref}
\usepackage[percent]{overpic}
\usepackage{overpic}
\usepackage{natbib}
\usepackage{multirow}
\usepackage{subfigure}
\usepackage{booktabs}  

\begin{document}

\newcommand{\bea}{\begin{eqnarray}}
\newcommand{\eea}{\end{eqnarray}}
\newcommand{\E}{\mathrm{E}}
\newcommand{\Var}{\mathrm{Var}}
\newcommand{\bra}[1]{\langle #1|}
\newcommand{\ket}[1]{|#1\rangle}
\newcommand{\braket}[2]{\langle #1|#2 \rangle}
\newcommand{\mean}[2]{\langle #1 #2 \rangle}
\newcommand{\be}{\begin{equation}}
\newcommand{\ee}{\end{equation}}	
\newcommand{\ba}{\begin{eqnarray}}
\newcommand{\ea}{\end{eqnarray}}
\newcommand{\SD}[1]{{\color{magenta}#1}}
\newcommand{\rem}[1]{{\sout{#1}}}
\newcommand{\alert}[1]{{\color{red} \uwave{#1}}}
\newcommand{\Y}[1]{\textcolor{blue}{#1}}
\newcommand{\R}[1]{\textcolor{red}{#1}}
\newcommand{\B}[1]{\textcolor{black}{#1}}
\newcommand{\C}[1]{\textcolor{cyan}{#1}}
\newcommand{\db}{\color{darkblue}}
\newcommand{\yx}[1]{\textcolor{cyan}{#1}}
\newcommand{\fan}[1]{\textcolor{red}{#1}}
\newcommand{\com}[1]{\textcolor{cyan}{#1}}
\newcommand{\intinfty}{\int_{-\infty}^{\infty}\!}
\newcommand{\Tr}{\mathop{\rm Tr}\nolimits}
\newcommand{\const}{\mathop{\rm const}\nolimits}

\title{Equation of state dependence of nonlinear mode-tide coupling in coalescing binary neutron stars}

\author{Yixiao Zhou  \altaffilmark{1} and
        Fan Zhang  \altaffilmark{2,3},
}

\altaffiltext{1}{Research School of Astronomy and Astrophysics, Australian National University, Canberra, ACT 2611, Australia}
\altaffiltext{2}{Gravitational Wave and Cosmology Laboratory, Department of Astronomy, Beijing Normal University, Beijing 100875, China}
\altaffiltext{3}{Department of Physics and Astronomy, West Virginia University, PO Box 6315, Morgantown, WV 26506, USA}

\begin{abstract}

Recently, an instability due to the nonlinear coupling of p-modes to g-modes in tidally deformed neutron stars in coalescing binaries has been studied in some detail. The result is significant because it could influence the inspiral and leave an imprint on the gravitational wave signal that depends on the neutron star equation of state (EOS). Because of its potential importance, the details of the instability should be further elucidated and its sensitivity to the EOS should be investigated. To this end, we carry out a numerical analysis with six representative EOSs for both static and non-static tides. We confirm that the absence of the p-g instability under static tides, as well as its return under non-static tides, is generic across EOSs, and further reveal a new contribution to it that becomes important for moderately high-order p-g pairs (previous studies concentrated on very high order modes), whose associated coupling strength can vary by factors of $\sim 10 - 100$ depending on the EOS. We find that, for stars with stiffer EOSs and smaller buoyancy frequencies, the instability onsets earlier in the inspiral and the unstable modes grow faster. These results suggest that the instability's impact on the gravitational wave signal might be sensitive to the neutron star EOS. To fully assess this prospect, future studies will need to investigate its saturation as a function of the EOS and the binary parameters.
\end{abstract}


\section{Introduction}

  Soon after the commissioning of the second generation gravitational wave (GW) detectors, which include the Advanced LIGO \citep{Harry:2010zz} and the Advanced Virgo \citep{aVIRGO:2012}, GWs from binary black holes were successfully detected \citep{2016PhRvL.116f1102A,2016PhRvL.116x1103A}. The next wave of excitement will likely come from neutron star coalescences (see \cite{2016ApJ...832L..21A} for an upper limit on event rates given the absence of detection from LIGO's first observing run), which would provide us with a new channel for probing the neutron star equation of state (EOS). Such a possibility has been examined in various studies. In particular, \cite{2008PhRvD..77b1502F,2010PhRvD..81l3016H,2012PhRvD..85l3007D,2013PhRvD..87d4001H,2013PhRvD..88d4042R} have developed schemes concentrating on the effects of the neutron stars' EOS-dependent tidal deformations on the gravitational waveform. A particularly interesting issue related to this line of investigation is whether tidally-driven instabilities can develop within the neutron stars, which would grow by draining energy from the orbital motion, keeping some as modal energy, and dissipating the rest as heat. Such instabilities, if they exist (1) and are sensitively dependent on the EOS (2), would impart signatures of the EOS onto the orbital and thus the gravitational waveform's phases, enhancing the GW detectors' ability to characterize the EOS.

  The question (1) regarding the existence of said instabilities has been examined by \cite{2013ApJ...769..121W} (WAB), \cite{2014ApJ...781...23V} (VZH) and \cite{2016ApJ...819..109W} (W2016). Previous study by \cite{2001ApJ...546..469W} demonstrated that three-mode coupling coefficients can become large when wave numbers of the two daughter modes become comparable. By considering such three-mode nonlinear couplings, and setting the daughter modes to be p-g pair with similar wavelength, WAB discovered a new non-resonant instability, with which a tidal force can quickly drive high order p- and g-modes to large amplitudes. Assuming a static tide, VZH then extended the calculation to include four mode couplings, by employing a novel volume-preserving transformation that greatly simplifies the computation. What they found is that a near-exact cancellation occurs between the three- and four-mode couplings, {which reduced the growth rate and implied that the instability cannot affect the inspiral significantly.} W2016 then further relaxed simplifying assumptions, allowing for volume-altering non-static tides (the stars become compressible under the influence of linear tides). The result is that the near-exact cancellation is undone, and the instability becomes important once again. In this paper, we confirm {that these conclusions for both the static and non-static cases are valid across EOSs, and also demonstrate the presence of an additional contributing factor to the instability}. We emphasize that our results differ from previous literature only in that we look at alternative additional terms that become important under different circumstances. Under the circumstances that are relevant to those works, the instabilities they found would still be the dominant contribution. We wish to demonstrate here that strong EOS dependence is present for at least some of the cases, so using the instabilities to study EOS would be a promising avenue, but a comprehensive dictionary of the instabilities for all possible mode pairs is far beyond the scope of the paper (and would be useful only when saturations, etc. are also thoroughly considered).

  In terms of observational consequences, the aforementioned studies {(see also \cite{2016PhRvD..94j3012E})} were mostly concerned with the detectability of GWs using matched-filtering techniques, supposing that the template waveforms do not account for the tidally-induced phase shifts correctly. Therefore, {the emphasis was on computing} whether the growth rate of the instabilities is in general large enough to make a significant alteration to templates necessary, and only a single fiducial EOS (SLy4) was invoked to provide concrete example numbers. 
Here, we will instead concentrate on the flip side of the story, and try to see if the timing for the onset of the instabilities (as calibrated to the orbital frequency), and thus the appearance of its alteration to the tidally induced phase shift, 
can be put to use and narrow down EOS possibilities. Our expectation that such a pursuit may be useful is born out of the observation that due to the very nature of instabilities, any small EOS-dependent variations in the related parameters would necessarily become amplified into ``diverging'' quantitative differences, even if broad qualitative features such as whether an instability appears is not sensitive to the EOS. We therefore focus on answering question (2): whether the nonlinear instability is sensitive to the EOSs. {To this end, we numerically implement the computational procedure developed by VZH and \cite{2012ApJ...751..136W}}, and compute the explicit p- and g-mode coupling strengths to the tide for six representative EOSs, in terms of their impacts on the g-mode frequencies. Our results show that they can easily differ by an order of magnitude, likely leading to observable effects. Physically, this result is not entirely surprising. Previous numerical study by \cite{2011MNRAS.418..427S} has suggested a sensitive dependence on the EOSs for nonlinear mode interactions in the post-merger hypermassive neutron stars (we concentrate instead on the pre-merger stage in this work). We caution however, that nonlinear instabilities are typically subjected to many complications that are beyond the scope of this paper. This omission is particularly acute when the unstable modes grow to large amplitudes. Therefore, an accurate determination of the instability window, and a quantitative description of the dissipation and mode-saturation effects are vitally important, before we can achieve a reliable EOS reading from GW signals using this type of instability.

  The remainder of the paper is organized as follows. Sect.~\ref{sec:EOS} is devoted to an introduction and comparison of the six typical EOSs. We demonstrate their properties by solving the Tolman-Oppenheimer-Volkoff (TOV) equations numerically. The main topic of Sect.~\ref{sec:Static} is the computation and comparison (across EOSs) of the mode-tide coupling strengths in the presence of a static tide. We then turn to non-static tides in Sect.~\ref{sec:Dynamic}, to re-evaluate the coupling constant and analyse the emerging instability. Finally, we conclude in Sect.~\ref{sec:Conclusion}.

\section{Equations of state} \label{sec:EOS}

  In this section, we briefly enumerate and compare the six EOSs included in our computations, providing their analytical forms where available and point to references for tabulated data. Due to the complicated and extensive nature of the field of study on neutron star EOSs, our introduction is necessarily cursory, and we refer the readers to review articles such as \cite{2012ARNPS..62..485L,2000PhR...328..237H} for more detailed discussions. 

\subsection{The six choices}
\subsubsection{SLy4}
\begin{figure}\begin{overpic}[width=0.9\columnwidth]{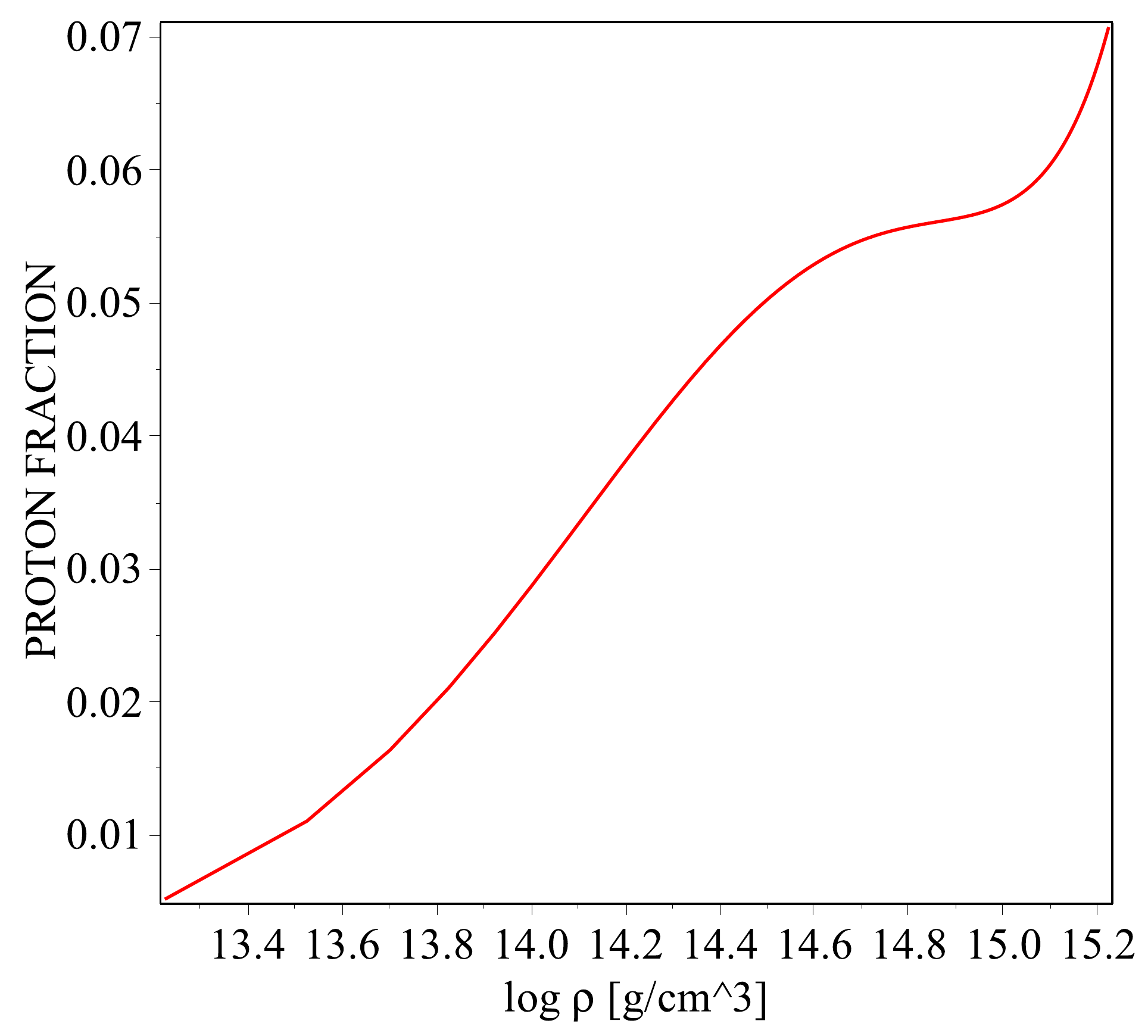}
\end{overpic}
\caption{The equilibrium relationship between the mass density $\rho$ and the proton fraction $Y_p$ for the SLy4 EOS, as computed by minimizing $\epsilon_{\rm tot}$ in Eq.~\eqref{eq:density}.}
\label{fig:SLy4Frac}
\end{figure}

The first EOS we include in our computations is the SLy4 \citep{1997NuPhA.627..710C,1998NuPhA.635..231C}, which is also adopted by WAB and W2016 as their fiducial example. This EOS is developed out of a refined Skyrme-like effective potential \citep{1959NucPh...9..615S}, originating from the shell-model description of the nuclei. SLy4 is simple enough such that analytical expressions for the EOS are readily available, aside from some parameters to be determined by fitting to experimental data. One starts with the total energy density
  \bea \label{eq:edensity}
  \epsilon_{\rm tot}(n_p,n_n,n_e)&=&n_pm_pc^2+n_nm_nc^2
 \notag \\
 && +n_bE_{\rm bind}(n_b,Y_p)+\epsilon(n_e)\,,
  \eea
where $n_p, n_n, n_e$ are proton, neutron, and electron number densities respectively, $n_b \equiv n_p+n_n$ is the baryon number density, and $Y_p$ is the proton fraction defined as $Y_p \equiv Z/A = n_p/n_b$. After imposing charge neutrality $n_p=n_e$, Eq.~\eqref{eq:edensity} simplifies into
\begin{equation}
\epsilon_{\rm tot}(n_b,Y_p)=n_b\bar{m}c^2+n_bE_{\rm bind}(n_b,Y_p)+\epsilon(n_p)\,,
\end{equation}
where $\bar{m}=(n_p m_p+n_n m_n)/n_b$ denotes the mean value of the nucleon mass. The term $E_{\rm bind}$ is the average binding energy per particle, {whose density functional is given in \cite{1997NuPhA.627..710C} Eq.~3.18. Parameters in this analytical expression are found in Tb.~1 of \cite{1998NuPhA.635..231C} for SLy4.} For the electron energy density $\epsilon(n_e)$, we note that the electron distribution is approximated by an ideal degenerate Fermi gas \citep{1983bhwd.book.....S}, hence 
\begin{equation} \label{eq:density}
\epsilon_{\rm tot}(n_b,Y_p)=n_b\bar{m}c^2+n_bE_{\rm bind}(n_b,Y_p)+\frac{\hslash c}{4\pi^2}(3\pi^2n_p)^{4/3}\,,
\end{equation}
where $\bar{m}c^2=938.91897\,\rm{MeV}$ and $\hslash c=197.32705\,\rm{MeV\cdot fm}$ are adopted from \cite{1997NuPhA.627..710C}. At a given baryon number density $n_b$, the equilibrium (called $\beta$-equilibrium in reference to the inverse $\beta$-decay) proton fraction $Y_p$ is the one that minimizes $\epsilon_{\rm tot}$. That is to say, equilibrium states depend only on $n_b$. For concreteness, we take $n_b$ from $0.01$ to $1.0\,{\rm fm}^{-3}$ in steps of $0.01$, and plot the corresponding equilibrium proton fraction values in Fig.~\ref{fig:SLy4Frac} (c.f.~Fig.~12 in \cite{1997NuPhA.627..710C}).
Subsequently, pressure $P$ and mass density $\rho$ can be obtained with the formula
\begin{equation}
\begin{aligned} \label{eq:Prho}
P(n_b,Y_p)&=n_b^2\frac{d(\epsilon_{\rm tot}/n_b)}{dn_b}\,,\\
\rho(n_b,Y_p)&=\frac{\epsilon_{\rm tot}}{c^2}\,.
\end{aligned}
\end{equation}
Substituting $n_b$ and the corresponding $Y_p$ into Eq.~\eqref{eq:Prho} then gives the $P-\rho$ relation, i.e. the equation of state.
  
\begin{figure}
\begin{overpic}[width=0.9\columnwidth]{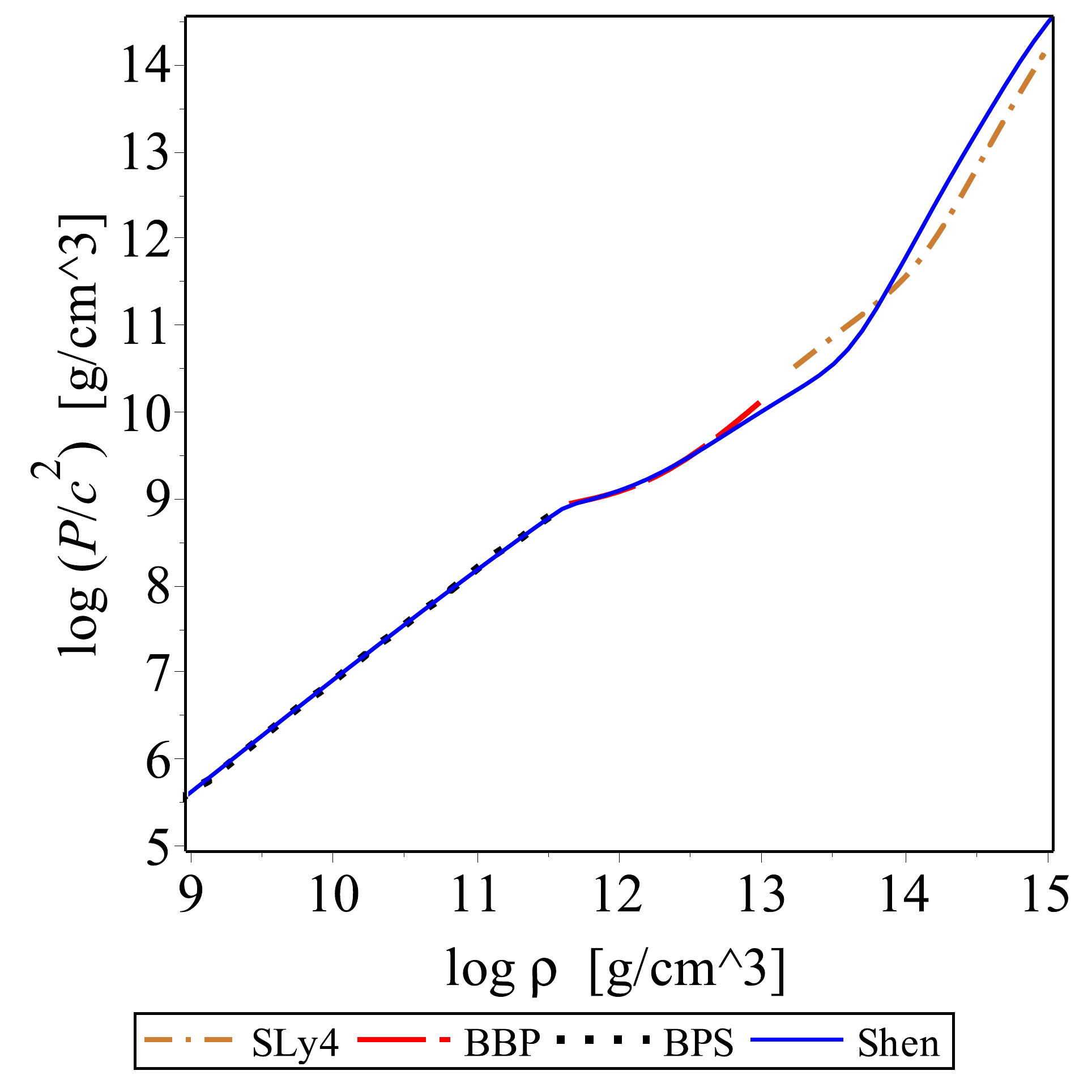}
\end{overpic}
\caption{{The SLy4, BBP, BPS and Shen EOSs. The SLy4 governs only the high density regime, while the segment with density below that of neutron drip ($\rho_{\rm drip}\approx 4\times10^{11}\rm g/cm^3$) is described by the BPS EOS. They are bridged by the BBP EOS. It turns out that neutron star properties (will be computed later) do not depend sensitively on the BPS or the BBP EOS, and we will simply refer the SLy4 + BBP + BPS combination as the SLy4 EOS in the following sections.}}
\label{fig:SLy4AllEOS}
\end{figure}

  The SLy4 EOS is applicable in the high-density regime ($10^{13} - 4\times10^{15}{\rm g/cm}^3$), and for below-neutron-drip densities, we supplement it with the 
Baym-Pethick-Sutherland (BPS) EOS \citep{1971ApJ...170..299B}. Moreover, for the connecting intermediate densities $(4\times10^{11} - 10^{13}{\rm g/cm}^3)$, we adopt the Baym-Bethe-Pethick (BBP) EOS \citep{1971NuPhA.175..225B}. The detailed tabulated data for both the BBP and the BPS EOSs are collected from \cite{1974ARA&A..12..167C}, and we plot all three aforementioned EOSs together in Fig.~\ref{fig:SLy4AllEOS}. We see that the transition between them is relatively smooth, without jumps at the seams that would signal potential inconsistencies. 

  
\subsubsection{Shen EOS}

  The Shen EOS \citep{1998NuPhA.637..435S,1998PThPh.100.1013S} is derived with a relativistic mean field (RMF) description of the nuclear matter, taking ingredients from quantum fields and the Hartree analysis for many-particle systems. It is a more sophisticated model than its non-relativistic counterparts, such as those based on the Skyrme force, for not only does it take into account the special relativistic effects, it also treats both nucleons and mesons.  
For a more comprehensive discussion regarding the RMF, please consult \cite{1990AnPhy.198..132G}. 
  
\begin{figure} 
\begin{overpic}[width=0.9\columnwidth]{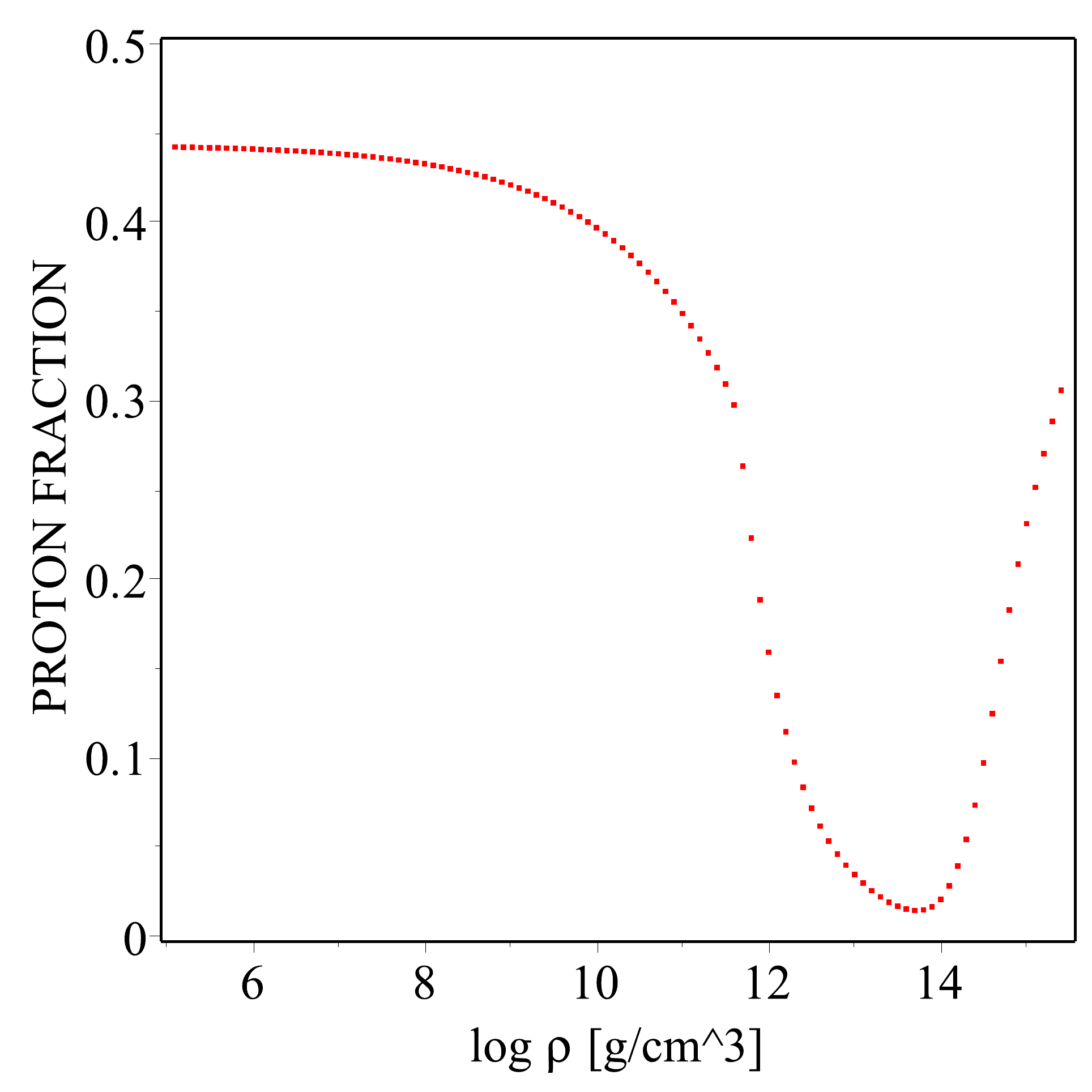}
\end{overpic}
\caption{The equilibrium relationship between the mass density $\rho$ and $Y_p$ for the Shen EOS (i.e., the $\beta$-equilibrium curve, c.f.~\cite{1998NuPhA.637..435S} Fig.~5).}
\label{fig:ShenFrac}
\end{figure}

In addition to being comparatively more thorough, Shen EOS also covers broad density and temperature ranges ($10^5<\rho <10^{15.5}{\rm g/cm}^3$; $0<T<100$MeV). For these reasons, it is widely adopted in supernova simulations and neutron star calculations. 
For example, \cite{2010CQGra..27k4106D} and \cite{2011MNRAS.418..427S} included the Shen EOS when studying black hole-neutron star mergers and the excitation of non-axisymmetric modes in the post-merger remnant, respectively. We will not go into any details about the derivation of this EOS, only pointing to its tabulated values on Shen's home page: \url{http://phy.nankai.edu.cn/grzy/shenhong/EOS/index.html}. 
We will use these data in the context of $T=0$, and note that although they contain both $\rho$ and $P$, only baryon contributions are accounted for. To add the influence of leptons and obtain a more complete EOS, we return to Eq.~\eqref{eq:density}, use the tabulated data to fill in $\epsilon_{{\rm bind}}$ at a given $\rho$, and then minimize $\epsilon_{{\rm tot}}$ to extract the proton fraction $Y_p$ at $\beta$-equilibrium, which is plotted in Fig.~\ref{fig:ShenFrac}. The first row of Eq.~\eqref{eq:Prho} then provides us with the full pressure including the lepton contributions. Repeating this procedure for various $\rho$ choices then results in the final EOS, which is depicted in Fig.~\ref{fig:SLy4AllEOS}. 

\subsubsection{Four APR equations of state} \label{sec:APREOS}

APR is the abbreviation of a series of four realistic EOSs developed by Akmal, Pandharipande and Ravenhall \citep{1998PhRvC..58.1804A}. All of them originate from nuclear physics and provide good fits to the two-nucleon scattering data. For convenience, we name them APR1 through 4. APR1 is the ``primary version'' of the APR EOSs, which is constructed from the Argonne $v_{18}$ potential that describes the interaction between two nucleons. On the basis of APR1, APR2 further considers relativistic boost effects while APR3 incorporates the Urbana model IX (UIX) describing interaction among three nucleons. Finally, APR4, the ``complete version'' of this series, include both the relativistic corrections and the three nucleon interaction potential UIX. The APR EOSs, especially APR4, is commonly used in neutron star simulations, for it appears to be compatible with astronomical observations (consult, for example, Fig.~8 in \cite{2013ApJ...773...11H}). Therefore, the APR EOSs are often referred to as being among the ``classical'' EOSs.

\begin{figure}
\begin{overpic}[width=0.9\columnwidth]{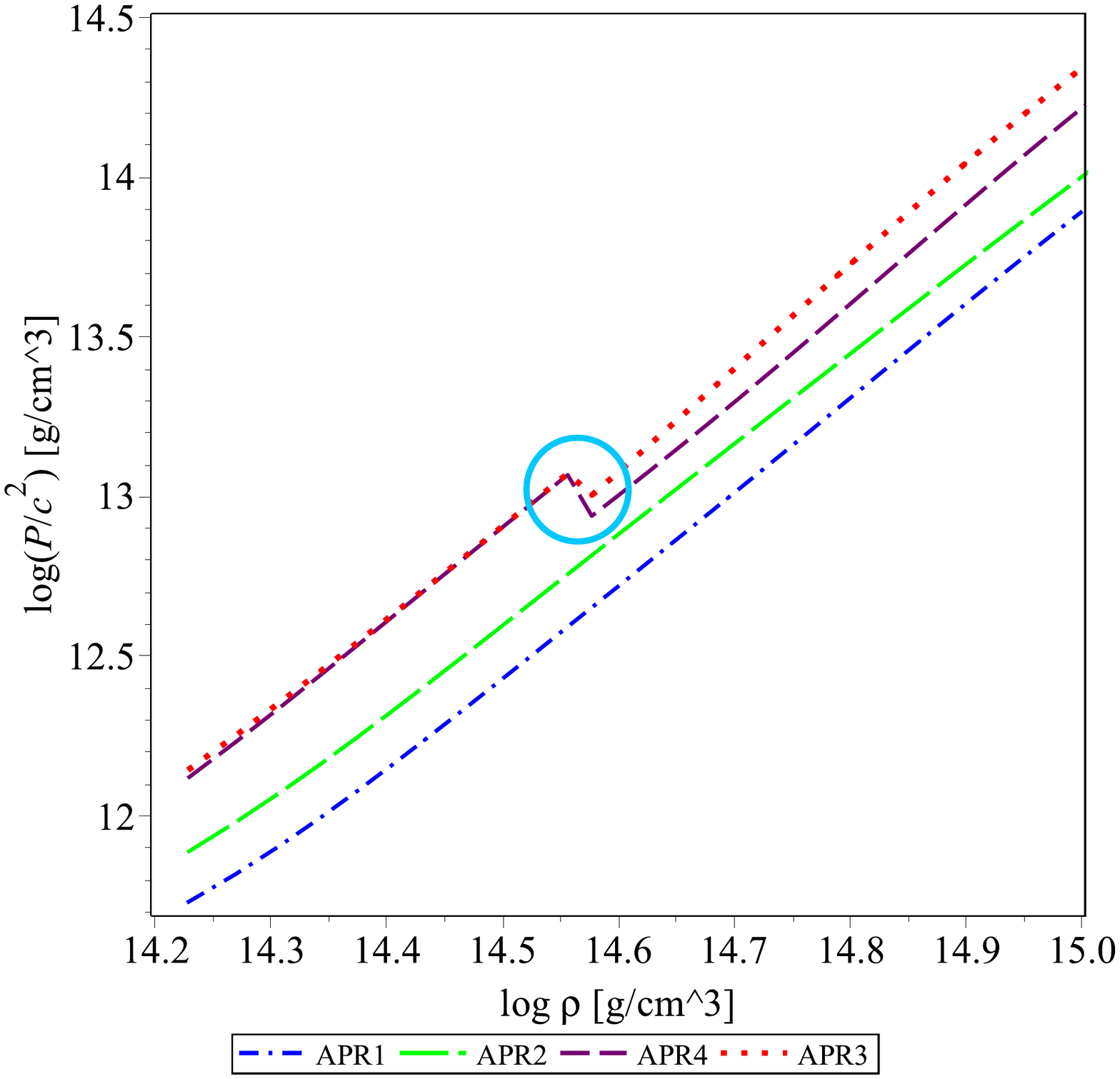}
\end{overpic}
\caption{Four APR EOSs. Note that the discontinuity of pressure in APR3 and APR4 (highlighted with cyan circles) represents a phase transition from normal neutron fluid to a phase with pion condensation, which occurs when one takes three-nucleon interactions into consideration.}
\label{fig:APREOS}
\end{figure}

The four APR EOSs are depicted in Fig.~\ref{fig:APREOS} (analytic expressions of the effective Hamiltonians and the corresponding parameters for the four EOSs can be found in Appendix A of \cite{1998PhRvC..58.1804A}). It is apparent that for any given mass density, APR1 has the lowest pressure while APR3 has the highest. This property is referred to as the ``softness'' (or ``stiffness'') of an EOS, describing the weakness (or strength) of the interaction between nuclear matter. Therefore, APR1 can be classified as a (relatively) soft EOS, whereas APR3 can be called stiff. Also noticeable is the discontinuity appearing in Fig.~\ref{fig:APREOS} for APR3 and APR4. This discontinuity represents a phase transition from normal neutron fluid to a phase with pion condensation\footnote{Transition from hadronic to quark matter is also disscussed in \cite{1998PhRvC..58.1804A}. Nevertheless, as stated in the same paper, this phenomena is not expected to happen inside a $1.4\rm{M_{\odot}}$ neutron star which is assumed for our study.}. Such transitions will not occur unless we consider the three-nucleon interactions, and are therefore not seen in APR1 and APR2. Finally, it is worth pointing out that APR EOSs cover only the high-density regime above 0.1 $\rm{fm^{-3}}$ (approximately the crust-core transition density). In our computation, we apply the FPS EOS \citep{1995NuPhA.584..675P} slightly below 0.1 $\rm{fm^{-3}}$ and continue it with the BBP EOS until the neutron drip density is reached. The BPS EOS is once again adopted at densities below neutron drip. We note that our choice of EOSs is consistent with \cite{1998PhRvC..58.1804A} (c.f.~Sect.~IV in that paper and \cite{1993PhRvL..70..379L}).
 
\subsection{Comparing the six equations of state} \label{sec:6EOS}

Although not large in number, our choice of the six typical (commonly invoked in literature) EOSs are extensive in the sense that they are derived with different techniques: non-relativistic effective potential for SLy4, relativistic mean field for Shen, variational calculation (also known as variational chain summation method or \textit{ab initio} calculation for many body system) for APR. Our typical six thus cover a majority of the approaches to modelling nuclear matter. We caution however that other models exist, including more exotic ones such as that of strange-quark (see \cite{1984PhRvD..30..272W}. It appears though that quark stars predicted with this theory is not consistent with the observation of a 1.97${\rm M}_\odot$ neutron star, the most massive one to date \citep{2010Natur.467.1081D,2013ApJ...773...11H}).
  
The six also cover a broad range of physical properties for their respective predicted neutron stars. In the spherically symmetric case, such properties can be computed by solving the   
TOV equations:  
\bea \label{eq:TOV}
  \frac{dm_{\ast}}{dr}&=&\frac{4\pi r^2\rho}{{\rm M}_{\odot}}\,,
\notag \\
  \frac{dP_1}{dr}&=&-\frac{G{\rm M}_{\odot}}{c^2}\frac{m_{\ast}}{r^2}(\rho+P_1)\left(1+\frac{4\pi r^3P_1}{m_{\ast}{\rm M}_{\odot}}\right)
\notag \\ &&\times  
  \left(1-\frac{G {\rm M}_{\odot}}{c^2}\frac{2m_{\ast}}{r}\right)^{-1}\,, \notag \\
  \frac{d\Phi_1}{dr}&=&\frac{G{\rm M}_{\odot}}{c^2}\frac{m_{\ast}}{r^2}\left(1+\frac{4\pi r^3P_1}{m_{\ast}{\rm M}_{\odot}}\right)\left(1-\frac{G{\rm M}_{\odot}}{c^2}\frac{2m_{\ast}}{r}\right)^{-1} \,,
\eea
where $m_{\ast}={\rm M}/{\rm M}_{\odot}$ is the dimensionless mass parameter, $P_1=P/c^2$ has the same dimension as $\rho$ (${\rm g/cm}^3$), and $\Phi_1=\Phi/c^2$ is the modified gravitational potential of a test particle with unit mass.
  
For our concrete numerical calculations, we fix the neutron star mass at the typical value of ${\rm M}\approx1.4{\rm M}_{\odot}$. To obtain the neutron star properties, we first specify an arbitrary central density $\rho_c$, before integrating the TOV equations until the surface of the star ($P_1=0$) is reached, at which stage we would have a value for the total mass $m_*$. Adjusting the $\rho_c$ value then allows us to drive $m_*$ towards $1.4$.
The radii and central densities thus obtained for different EOSs are tabulated in Tb.~\ref{tb:NSProperty}. It is clear that a softer EOS has a denser core and a smaller size.
   
\begin{figure}
\begin{overpic}[width=0.9\columnwidth]{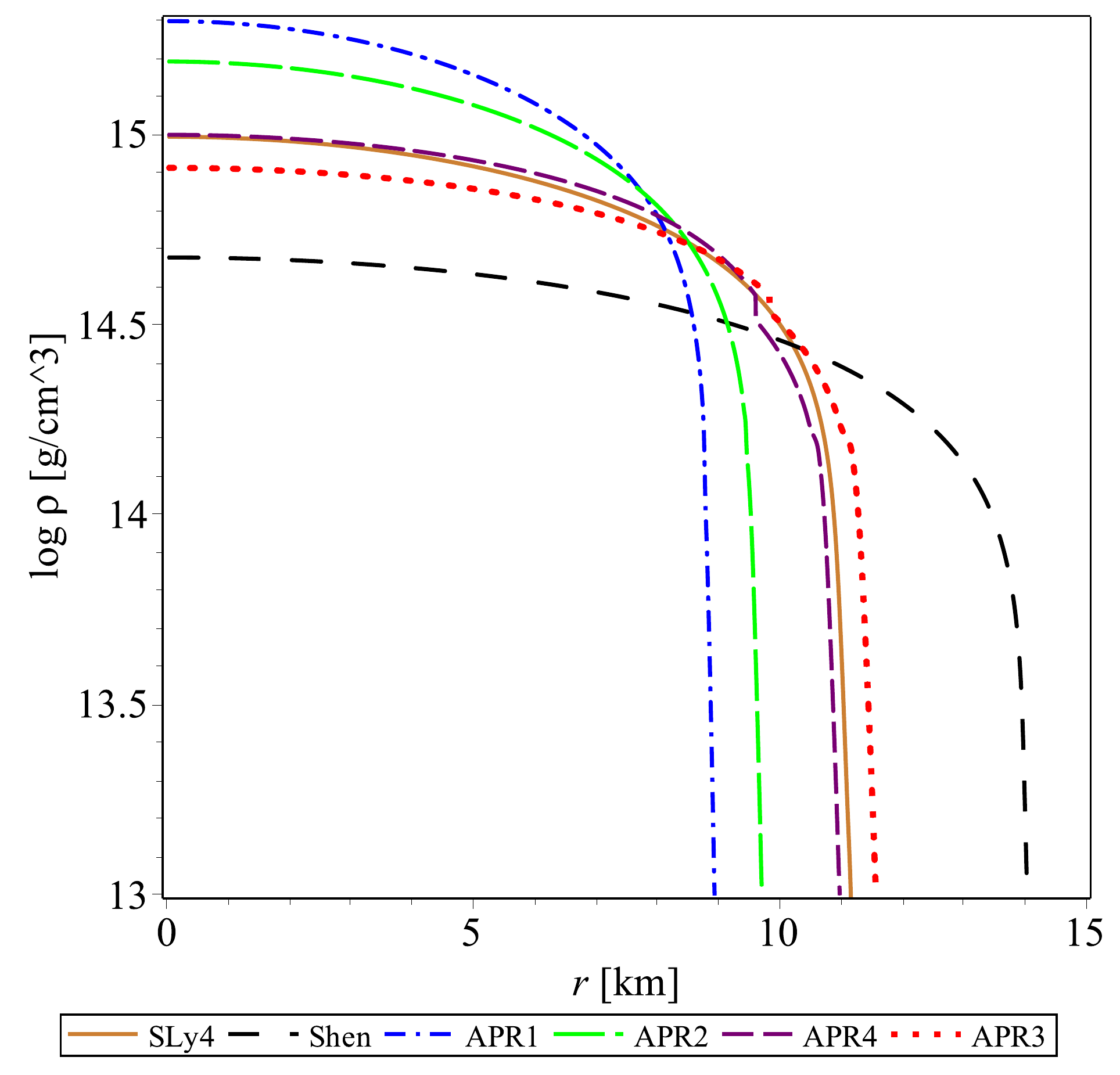}
\end{overpic}
\caption{The mass density profiles within neutron stars predicted by the six typical EOSs. All of them share a common feature, i.e., density varies slowly in the core region and drops swiftly near the surface. For APR3 and APR4, the leap of density inside the star results from the phase transition (see Fig.~\ref{fig:APREOS}).}
\label{fig:rrho}
\end{figure}
\begin{figure}
\begin{overpic}[width=0.9\columnwidth]{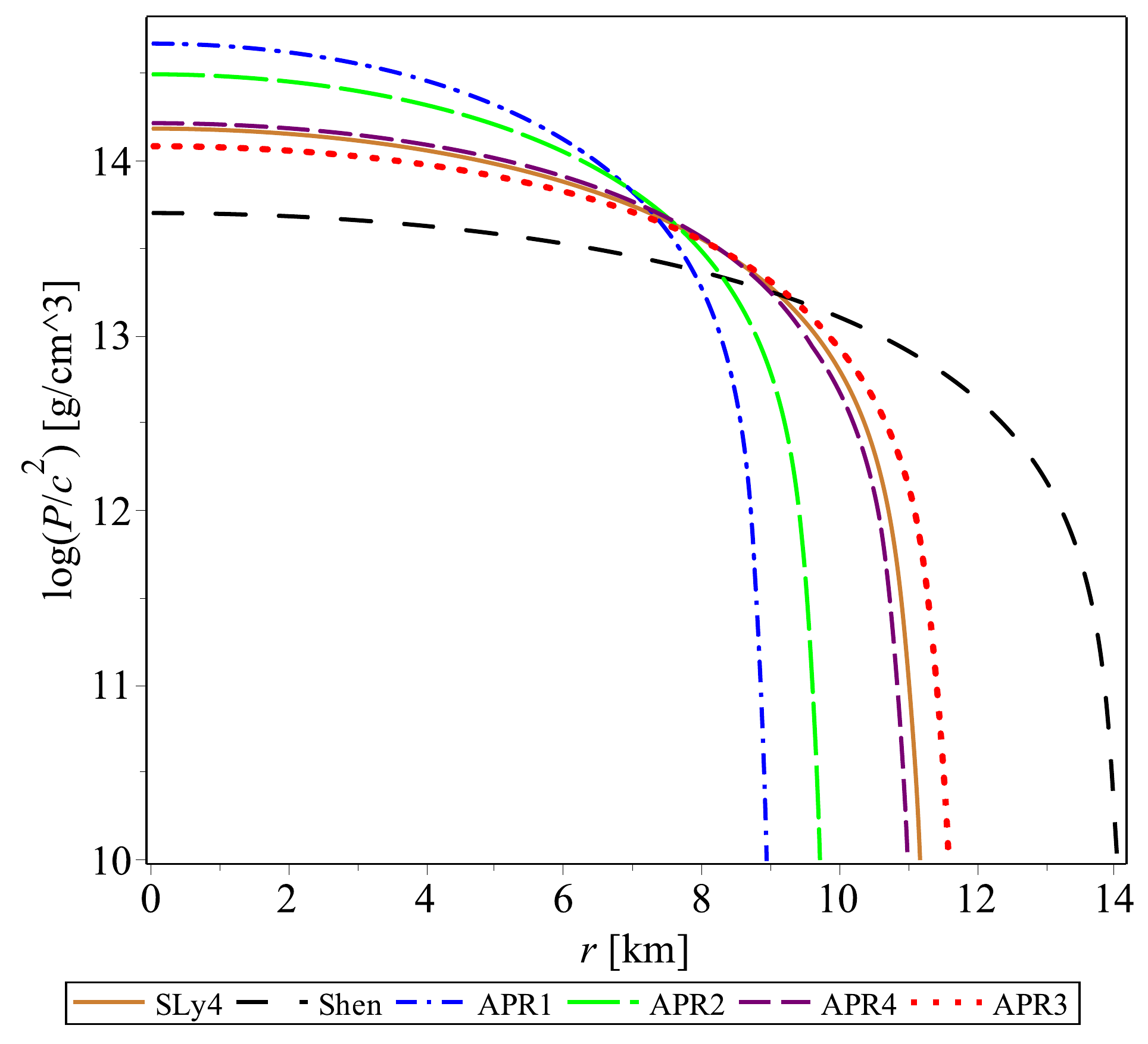}
\end{overpic}
\caption{The pressure profiles within neutron stars predicted by the six typical EOSs. For nuclear matter, softer EOSs imply weaker interactions at any given density. For neutron stars, softer EOSs lead to greater pressure in the core for a fixed total mass. \label{fig:rP}}
\end{figure}

\begin{table}[b!]
\centering
\caption{The radii and central densities of neutron stars with mass $1.4{\rm M}_{\odot}$, according to different EOSs. \label{tb:NSProperty}}
{\begin{tabular*}{1.03\columnwidth}{@{\extracolsep{\fill}}ccccccc}
\toprule[2pt]
  EOS & SLy4 & Shen & APR1 & APR2 & APR3 & APR4 \\ 
\midrule[1pt]
  $\mathcal{R}$ (km) & 11.663 & 14.921 & 9.205 & 10.051 & 12.132 & 11.461 \\ 

  $\log\rho_c ({\rm g/cm}^3)$  & 14.995 & 14.677 & 15.300 & 15.193 & 14.913 & 15.000 \\ 
\bottomrule[2pt]
\end{tabular*}}
\end{table}

Figs.~\ref{fig:rrho} and ~\ref{fig:rP} further depict the detailed distributions of density and pressure inside the star. Since the same overall mass is shared across all EOSs, we immediately see that softer EOSs with more matter concentrated in the core predict more compact stars. From these figures, we can also assess the stiffness of the SLy4 and Shen EOSs. The six EOSs, ordered from stiff to soft, are Shen, APR3, SLy4, APR4, APR2 and APR1, forming a rather evenly spaced sequence with no one being redundant.

\section{Static Tide} \label{sec:Static}

\subsection{The road to mode-tide coupling strength} \label{sec:roadtoMTCS}

We begin by presenting the expressions for the {mode-tide coupling strength (MTCS)}, by which we mean the terms driving the g-mode frequency shift due to its nonlinear coupling to a p-mode and the tide. As we will see later, the value of this frequency shift tells us when and how g-modes get driven by the tide into exponential growth, and is therefore the most conspicuous manifestation of the mode-tide coupling. The detailed derivations leading to the shifts are involved and tedious, but a much simpler expression exists {for a static} (time-independent) tide, derived by VZH during their stability studies. We will briefly review their approach, following their steps even though we will use their final result for a different purpose. 

Starting with an isolated neutron star (not subjected to any tidal forces), its oscillation modes can be 
acquired by solving asteroseismology equations assuming separability of variables. Specifically, the Lagrangian displacement of fluid\footnote{As with WAB, VZH and W2016, we assume completely fluid neutron stars. The solid crust does have an impact on core g-modes. However, its effect is verified to be small (please consult \cite{1992ApJ...395..240R} Sect.~5.2, see also footnote 6 in WAB and the introduction section in VZH).} elements (as the difference between the actual Eulerian and initial locations) due to a mode labelled $a$ can be written as 
\begin{equation} \label{eq:sepaofxi}
\begin{aligned}
&\vec{\xi}_a=\xi_r(r)Y_{l_am_a}(\theta,\phi)\hat{r}\\
&+\xi_h(r)\left(\partial_\theta Y_{l_am_a}(\theta,\phi)\hat{\theta}+\frac{1}{\sin\theta}\partial_\phi Y_{l_am_a}(\theta,\phi)\hat{\phi}\right),
\end{aligned}
\end{equation}
where $\xi_r$ and $\xi_h$ are the radial functions, and $Y_{l_am_a}(\theta,\phi)$ are the spherical harmonic functions with $l_a,m_a$ being the angular quantum numbers. These modes form an orthogonal basis, satisfying the relation
\begin{equation} \label{eq:Orthogonality}
\begin{aligned}
\int d^3\vec{x}\rho\vec{\xi_a^*}\cdot \vec{\xi_b}=\frac{E_0}{\omega_a^2}\delta_{ab},
\end{aligned}
\end{equation}
where $E_0\equiv GM^2/\mathcal{R}$ is a normalization constant that corresponds to the energy in a mode of unit amplitude, and $\omega_a$ is the angular eigenfrequency of mode $a$. One can then decompose any generic displacement $\vec{\gamma}$ to the fluid in the star (can be from the tide, excitation of modes or any other source) as 
\begin{equation} \label{eq:modeexpan}
\begin{aligned}
\vec{\gamma} = \sum \limits_a \chi_a\vec{\xi_a}\,.
\end{aligned}
\end{equation}
When a collection of these basis modes are excited, they will evolve according to their contributions to the total potential energy of the star (through the usual Lagrangian mechanics derivations), and will in general couple nonlinearly with coupling constants 
\begin{equation} \label{eq:MTCSdefinition}
\begin{aligned}
\kappa_{abc}&=-\frac{1}{2E_0}\int d^3\vec{x}\rho f_3\left(\vec{\xi_a},\vec{\xi_b},\vec{\xi_c}\right)\,,\\
\kappa_{abcd}&=-\frac{1}{6E_0}\int d^3\vec{x}\rho f_4\left(\vec{\xi_a},\vec{\xi_b},\vec{\xi_c},\vec{\xi_d}\right)\,,
\end{aligned}
\end{equation}
where $f_n$ represent the form with which the modes contribute to the potential energy at the $n$-th order of their amplitudes. 

Now let's introduce a tidal influence, taking for a prototype twin neutron stars in a binary, i.e., two stars with the same mass ${\rm M}$, radius $\mathcal{R}$ and are separated from each other by a distance $A$. For further simplification (to be relaxed later in Sect.~\ref{sec:Dynamic}), we ignore the more dynamical effects of the orbital motion, and view the two stars as being in rest and frozen in place in terms of their centers of mass. The static (time-independent) tide on one of the stars resulting from the gravitational field emanating from the companion star is described by the tidal potential $\epsilon U$, which is obtained from basic celestial mechanics. Keeping to the leading (quadrupolar) order in a spherical harmonics expansion, we have that 
\begin{equation} \label{eq:equitide}
  \begin{aligned}
  U\approx -\omega_0^2r^2P_2(\cos\theta),
  \end{aligned}
  \end{equation}
  with $P_2(\cos\theta)$ being the $l=2$ Legendre polynomial, $\epsilon\equiv \mathcal{R}^3/A^3$ being the tidal strength and $\omega_0\equiv\sqrt{GM/\mathcal{R}^3}$ the characteristic dynamical frequency. This $\epsilon U$ enters into the potential energy and changes the equations of motion for the fluid elements, thus causing a change to the modes. Namely the original modes of the isolated neutron star are perturbed in a tidally deformed star. Applying the usual perturbation theory (similar to the familiar one from quantum mechanics), we expect that the modal frequencies will be shifted. Indeed, detailed calculation in VZH shows that 
\begin{equation} \label{eq:perturbedgmode1}
\begin{aligned}
\frac{\omega_-^2}{\omega_g^2}= &1-\epsilon\left(U_{\bar{g}g}+\sum \limits_a  2\kappa_{a\bar{g}g}\chi_a^{(1)}\right)\\
&-\epsilon^2\sum \limits_{a,b}\left(2\kappa_{a\bar{g}g}\chi_a^{(2)}+3\kappa_{ab\bar{g}g}\chi_a^{(1)}\chi_b^{(1)}\right)\\
&-\epsilon^2\frac{\omega_p^2}{\omega_p^2-\omega_g^2}\left\vert U_{\bar{p}g}+\sum \limits_a2\kappa_{a\bar{p}g}\chi_a^{(1)}\right\vert^2 + \mathcal{O}(\epsilon^3)\,,
\end{aligned}
\end{equation}
where $\omega_-$ is the perturbed g-mode frequency (the potential instability lies in the perturbed high order g-modes that have small initial $\omega_g^2$ to start with (VZH), so such modes are the focus here) when a pair of daughter p- and g-modes nonlinearly couple to the tide. The symbol $\bar{a}$ indexes the complex conjugation of the basis vector of mode \textit{a}, and the reality condition demands that the coefficient to $\vec{\xi}_a$ is related to its complex conjugate counterpart through a parity factor $(-1)^{m_a}$.

The quantities $\chi_a^{(i)}$ and $U_{ab}$ are defined by
\bea \label{eq:defUab}
\vec{\chi} &=& \sum_a \left(\epsilon \chi^{(1)}_a +\epsilon^2 \chi_a^{(2)}\right)\vec{\xi}_a, \notag \\
U_{ab}&=&-\frac{1}{E_0}\int d^3\vec{x}\rho \vec{\xi_a}\cdot \left(\vec{\xi_b}\cdot \nabla\right)\nabla U\,,
\eea 
where the tidal deformation $\vec{\chi}$ is the static response of the neutron star to the tide. Worth noting is the appearance of three- and four-mode coupling constants in Eq.~\eqref{eq:perturbedgmode1}. They are obviously there to account for the coupling between the neutron star eigen-modes and the tidal deformation (as the primary perturbation), as their contribution vanishes when $\epsilon=0$. Although we have decomposed $\vec{\chi}$ into modal basis, and thus the overall coupling to tide into modal pieces, we can define a {``vector potential"}
\bea \label{eq:PotVec}
U_a = -\frac{1}{E_0} \int d^3 x \rho  \,\vec{\xi}_a^{*} \cdot \nabla U\,,
\eea 
and carry out resummations such as $\sum_c \kappa_{ggc}U_c$ to reassemble quantities into forms that are more directly identifiable as being ``tidal'' and more economical in notation (see e.g. the left hand side of Eq.~\eqref{eq:MTCS} below). 
With regard to stability, previous investigations (WAB, VZH) have proven that $\sum_a \kappa_{apg}\chi_a^{(1)}\sim \omega_p/\omega_g$, which implies that such couplings can be large for higher order p-g pairs (i.e. a high frequency p-mode and a low frequency g-mode), and thus drive the left-hand-side of Eq.~\eqref{eq:perturbedgmode1} negative (barring any cancellations). This is the non-resonant p-mode g-mode instability discovered by WAB.

Before proceeding further, we note that a judicious choice for the definition of the displacements would likely simplify computations. Instead of defining them as the difference between the Eulerian coordinates and the initial coordinates in the isolated neutron star, the alternative of using an initial coordinate system more suited to the tidally deformed star appears to make sense. Such an approach is adopted by VZH who developed a novel technique called the volume preserving transformation (VPT, briefly reviewed in Appendix \ref{app:VPT}). This transformation maps a tidally deformed star into a radially stretched spherical star of equal volume. By comparing the potential energy in the two coordinate systems, they arrive at the transformation rules 
 \begin{equation} \label{eq:match1}
\begin{aligned}
U_{\bar{a}b}+\sum \limits_c  2\kappa_{\bar{a}bc}\chi_c^{(1)}=-\left(J_{\bar{a}b}^{(1)}+J_{b\bar{a}}^{(1)}\right),
\end{aligned}
\end{equation} 
and 
\begin{equation} \label{eq:match2}
\begin{aligned}
&\sum \limits_{c,d}\left(2\kappa_{\bar{a}bc}\chi_c^{(2)}+3\kappa_{\bar{a}bcd}\chi_c^{(1)}\chi_d^{(1)}\right)\\
&=-\sum \limits_{c}\left(J_{c\bar{a}}^{(1)}J_{cb}^{(1)}+J_{\bar{a}b}^{(2)}+J_{b\bar{a}}^{(2)}-2\kappa_{\bar{a}bc}V_c-V_{\bar{a}b}\right),
\end{aligned}
\end{equation}
where the definitions of $V_a$ and $V_{ab}$ are formally the same as those of $U_a$ and $U_{ab}$, but with the derivatives taken against the new coordinates (so $V_a$ is purely radial), and $J_{ab}^{(i)}$ is the i-th order (in $\epsilon$) Jacobian of the VPT. 

Given the rules \eqref{eq:match1}, \eqref{eq:match2} and the fact that $\omega_p^2\gg \omega_g^2$ for high-order p- and g-mode, the expression for the perturbed g-mode frequency becomes 
\begin{equation} \label{eq:perturbedgmode2}
\begin{aligned}
&\frac{\omega_-^2}{\omega_g^2}\approx1+2\epsilon J_{\bar{g}g}^{(1)}-\epsilon^2\left(J_{\bar{p}g}^{(1)}+J_{g\bar{p}}^{(1)}\right)\left(J_{pg}^{(1)}+J_{gp}^{(1)}\right)\\
&+\epsilon^2\sum \limits_{c=\lbrace p,g\rbrace}\left(J_{c\bar{g}}^{(1)}J_{cg}^{(1)}+J_{\bar{g}g}^{(2)}+J_{g\bar{g}}^{(2)}-2\kappa_{\bar{g}gc}V_c-V_{\bar{g}g}\right)\\
&\approx 1\pm\bigg\{ 2\epsilon J_{gg}^{(1)}-\epsilon^2\left(2\kappa_{gg\sigma}+V_{gg}\right)\\
&+\epsilon^2\left[\left(J_{gg}^{(1)}\right)^2-\left(J_{gp}^{(1)}\right)^2-2J_{pg}^{(1)}J_{gp}^{(1)}+2J_{gg}^{(2)}\right]\bigg\},
\end{aligned}
\end{equation}
where $\kappa_{ab\sigma}\equiv \sum_c\kappa_{abc}V_c$. The sign in the third line of Eq.~\eqref{eq:perturbedgmode2} is determined by $m_p$ and $m_g$. A plus sign corresponds to even $m_{p}$ and $m_{g}$\footnote{The p-g pair must share the same parity in order to satisfy the selection rule.}, while odd $m_{p}$ and $m_{g}$ give the minus sign.
The right-hand side of Eq.~\eqref{eq:perturbedgmode2} represents the impact of the tide on the g-mode that is coupled to it.
Comparing Eq.~\eqref{eq:perturbedgmode2} with Eq.~\eqref{eq:perturbedgmode1}, we see that after the VPT, the explicit four-mode coupling term drops out, greatly simplifying the derivation. We also note that the last line in Eq.~\eqref{eq:perturbedgmode2} is in fact much smaller than the rest of the terms on the right-hand side (an explanation is provided in Appendix \ref{app:Jaco}), so we can drop it to obtain
\begin{equation} \label{eq:perturbedgmode3}
\begin{aligned}
\frac{\omega_-^2}{\omega_g^2}\approx 1\pm\left[2\epsilon J_{gg}^{(1)}-\epsilon^2\left(2\kappa_{gg\sigma}+V_{gg}\right)\right]\,.
\end{aligned}
\end{equation}
Since only $m_{p,g}$ choices that lead to potential instabilities are of interest to us, we specialize to the cases where the minus sign is taken above. 
The remaining Jacobian contribution to Eq.~\eqref{eq:perturbedgmode3} is given by (VZH Eq.~89) 
\begin{equation} \label{eq:firstJ}
\begin{aligned}
\epsilon J_{gg}^{(1)}=-\epsilon\frac{\omega_g^2}{E_0}I_{gg\chi^{(1)}}\,,
\end{aligned}
\end{equation}
and the integral $I_{gg\chi^{(1)}}$ is given by Eq.~\eqref{eq:integral} (VZH Eq.~81). With static tide, this Jacobian term appears at $\mathcal{O}(\epsilon)$ and is non-negligible when the two neutron stars are far apart. However, it makes a much smaller contribution (see Tb.~\ref{tb:staticMTCS} below) in the more interesting late-inspiral regime. Therefore, although we compute its values (the details are provided in Appendix \ref{app:Jaco}) for completeness, we will exclude it from the definition of the MTCS. Instead, we will refer to the absolute value of $\epsilon^2(2\kappa_{gg\sigma}+V_{gg})$ as the MTCS, and note that its detailed expression is provided by VZH Eq.~99, which we reproduce here: 
\begin{equation} \label{eq:MTCS}
\begin{aligned}
& {\rm {MTCS}} \equiv \epsilon^2 \vert 2\kappa_{gg\sigma} + V_{gg}\vert =
\\ &
-\frac{1}{E_0}\int dr \left\{ r^2P\left[\Gamma_1(\Gamma_1+1)+\left(\frac{\partial\Gamma_1}{\partial\ln \rho}\right)_s\right]
\right. \\ &\left.
\times (\nabla\cdot\vec{\sigma})(\nabla\cdot\vec{g})_r^2
\right. \\ &\left.
-4r\sigma_r\Gamma_1P(\nabla\cdot\vec{g})_r^2
-\rho \mathfrak{g}r^3g_r^2\frac{d^2}{dr^2}\left(\frac{\sigma_r}{r}\right)
\right. \\ &\left.
-2r^2 \left[ \Lambda_g^2\omega_g^2\rho rg_h^2+2g_r\Gamma_1P(\nabla\cdot\vec{g})_r\right] \frac{d}{dr}\left( \frac{\sigma_r}{r} \right)
\right. \\ &\left.
+\left[-\rho \mathfrak{g}^2r\frac{d}{dr}\left(\frac{\sigma_r}{\mathfrak{g}}\right) + \rho r\epsilon^2\frac{dV}{dr}\right]
\right. \\ &\left.
\times \left[2rg_r(\nabla\cdot\vec{g})_r+g_r^2\frac{d\ln \rho}{d\ln r}\right] \right\}\,,
\end{aligned}
\end{equation}
where $\Lambda_g^2\equiv l_g(l_g+1)$. We will discuss the quantities appearing in Eq.~\eqref{eq:MTCS} in more details below, but mention that the derivation of this equation has invoked the Cowling approximation, i.e., it neglects the Eulerian perturbation to the gravitational potential (denoted $\Phi^{\prime}$). The Cowling approximation is reasonable for high order modes because when the radial and angular quantum numbers $n$ and $l$ are large, $\Phi^{\prime}$ is very small as compared to the Eulerian perturbations to the density $\rho$ and the pressure $P$ (see \cite{CDBook} Ch.~5.2). 

\subsection{The ingredients in the {\rm MTCS}} \label{sec:ingredMTCS}
As Eq.~\eqref{eq:MTCS} is central to our analysis, we devote this section to explaining the quantities appearing in it and demonstrating how to compute them. 

\subsubsection{$\mathfrak{g}$ and $\Gamma_1$}
The symbol $\mathfrak{g}$ refers to the local gravitational acceleration. Following VZH, we define $\mathfrak{g} \equiv d\Phi/dr$ rather than the usual $\mathfrak{g}=-\nabla \Phi$. $\Gamma_1\equiv (\partial\ln P/\partial\ln \rho)_{s}$ is the adiabatic index, which is in principal not the same as the polytropic exponent $\Gamma\equiv d\ln P/d\ln \rho$ in the equilibrium state. The appearance of $\Gamma_1$ in Eq.~\eqref{eq:MTCS} implies that only adiabatic oscillation is discussed in this paper. That is, we assume that the system is thermally isolated and the entropy does not change ($\Delta s=0$) throughout our discussion. $\mathfrak{g}$ can  be obtained straightforwardly by solving the TOV equations \eqref{eq:TOV} while $\Gamma_1$ must be derived from the EOS itself.
  
\subsubsection{$g_r$ and $g_h$} \label{sec:grgh}

The functions $g_r$ and $g_h$ are the radial and horizontal components of the g-mode eigenfunction (see Eq.~\eqref{eq:sepaofxi}). They are govened by the equations of oscillation (\cite{1989nos..book.....U} Eq.~13.1-13.3)
\begin{equation} \label{eq:governingequation_gr}
\begin{aligned}
&\frac{1}{r^2}\frac{d}{dr}(r^2g_r)-\frac{\mathfrak{g}}{c_s^2}g_r+\left(1-\frac{l_g(l_g+1)c_s^2}{r^2\omega_g^2}\right)\frac{P^{\prime}}{\rho c_s^2} \\
&=\frac{l_g(l_g+1)}{r^2\omega_g^2}\Phi^{\prime}, \\
&\frac{1}{\rho}\frac{dP^{\prime}}{dr}+\frac{\mathfrak{g}}{\rho c_s^2}P^{\prime}+(N^2-\omega_g^2)g_r=-\frac{d\Phi^{\prime}}{dr}, \\
&\frac{1}{r^2}\frac{d}{dr}\left(r^2\frac{d\Phi^{\prime}}{dr}\right)-\frac{l_g(l_g+1)}{r^2}\Phi^{\prime}=4\pi G\rho \left(\frac{P^{\prime}}{\rho c_s^2}+\frac{N^2}{\mathfrak{g}}g_r\right).
\end{aligned}
\end{equation}
\begin{equation} \label{eq:governingequation_gh}
\begin{aligned}
g_h=\frac{1}{r\omega_g^2}\left(\frac{P^{\prime}}{\rho}+\Phi^{\prime}\right),
\end{aligned}
\end{equation}
where $c_s\equiv \sqrt{\Gamma_1P/\rho}$ is the adiabatic sound speed, with a value of $c_s\sim 0.1c$ in the neutron star centre and must not exceed the speed of light $c$ anywhere by causality. $P^{\prime}$ and $\Phi^{\prime}$ are the Eulerian perturbations to the pressure and the gravitational potential due to the g-mode, respectively, and $N$ is called the buoyancy frequency (also called the Brunt-V\"ais\"al\"a frequency, please see further discussions in Sect.~\ref{sec:buoyancy_fre}).
Eq.~\eqref{eq:governingequation_gr} is the standard equation system describing the non-radial oscillations of the star, originating from the continuity equations, the hydrostatic equations for fluids and the Poisson equation, and is simplified to this form under the assumption of adiabatic oscillation. We solve Eqs.~\eqref{eq:governingequation_gr} and \eqref{eq:governingequation_gh} for $g_r, g_h$ and the eigen-frequency $\omega_g$ numerically, using the Aarhus adiabatic oscillation package (\textsc{adipls}, please consult \cite{2008Ap&SS.316..113C} for a thorough introduction to the \textsc{adipls} and its usage) with the boundary conditions of
\begin{equation} \label{eq:inner_boundary}
\begin{aligned}
g_r&=l_gg_h, \\
\frac{d\Phi^{\prime}}{dr}&=\frac{l_g}{r}\Phi^{\prime},
\end{aligned}
\end{equation}
at the center ($r=0$) and
\begin{equation} \label{eq:surface_boundary}
\begin{aligned}
\delta P&=0, \\
\frac{d\Phi^{\prime}}{dr}&=-\frac{l_g+1}{r}\Phi^{\prime},
\end{aligned}
\end{equation}
on the surface ($\delta P$ represents the Lagrangian perturbation to pressure).

\begin{figure}
\begin{overpic}[width=0.9\columnwidth]{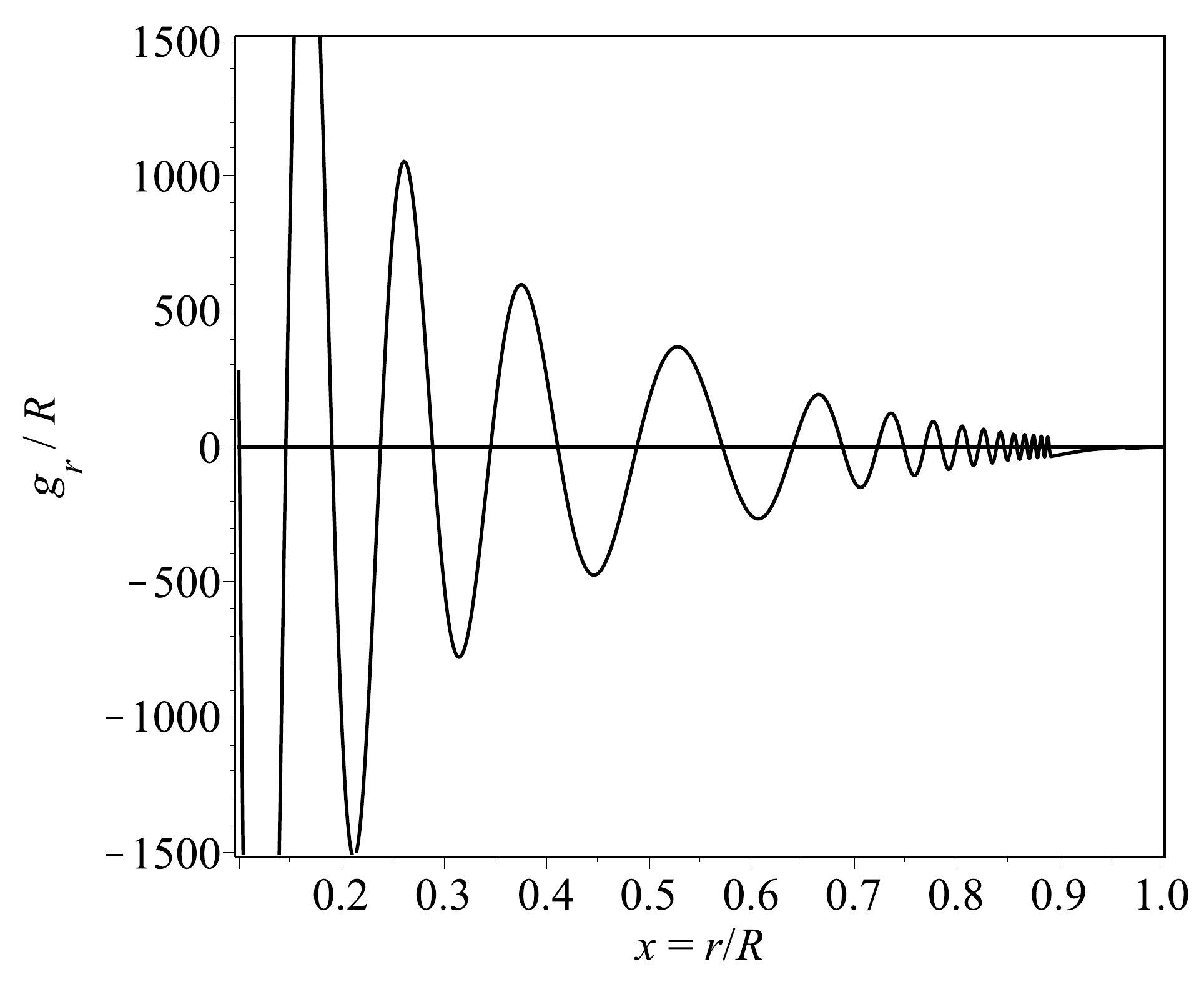}
\end{overpic}
\caption{The scaled radial function for a $l_g=4$, $n=32$ g-mode with frequency $f_g\approx 2.7$ Hz. $g_r$, as well as $f_g$, are computed with the \textsc{adipls} package within the SLy4 model.}
\label{fig:SLy4_eigenf}
\end{figure}

  {An example} $l_g=4$, $n=32$ g-mode under the SLy4 model is demonstrated in Fig.~\ref{fig:SLy4_eigenf}. It is obvious that fluid elements oscillate severely in the deep interior of the star; in contrast, the oscillations become relatively subdued near the surface. This is typical of g-modes (see, e.g., Fig.~5.10 in \cite{CDBook}). Additionally, we clarify that in order to be consistent with WAB, VZH and W2016, the normalization rule for $g_r$ and $g_h$ (equivalently, the definition of the normalization constant $E_0$) in this paper is given by Eq.~\eqref{eq:Orthogonality}, rather than Eq.~36 in \cite{2008Ap&SS.316..113C}.

Another associated constituent appearing in Eq.~\eqref{eq:MTCS} is $(\nabla\cdot\vec{g})_r$, the radial component of the divergence of the g-mode displacement vector. Its expression is given by
\begin{equation} \label{eq:div_g}
\begin{aligned}
(\nabla\cdot\vec{g})_r=\frac{\rho}{\Gamma_1 P}(\mathfrak{g}g_r-\omega_g^2rg_h+\Phi^\prime),
\end{aligned}
\end{equation}
which is actually equivalent to the first equation in \eqref{eq:governingequation_gr}\footnote{To find the connection between Eqs.~\eqref{eq:governingequation_gr} and \eqref{eq:div_g}, one could substitute the definition of divergence (VZH Eq.~96) and Eq.~\eqref{eq:governingequation_gh} into Eq.~\eqref{eq:div_g}.}. In terms of numerical computations, it is safe to ignore the last two terms in Eq.~\eqref{eq:div_g} because neither of them is comparable with the first one throughout the entire neutron star. For our neutron star models, $\Phi^{\prime}$ is a mere one thousand in magnitude as compared to $\mathfrak{g}g_r$, and the second term is even smaller than $\Phi^{\prime}$ for low frequency g-modes.

\subsubsection{N} \label{sec:buoyancy_fre}

The buoyancy frequency $N$, although not present explicitly in Eq.~\eqref{eq:MTCS}, does have an indirect influence on the MTCS through its strong impacts on $g_r$ and $g_h$ (Eq.~\eqref{eq:governingequation_gr}). Thus, we devote this subsection to the computation of $N$. The full definition for the buoyancy frequency is 
\bea \label{eq:N_def}
N^2\equiv \mathfrak{g}^2\left(\frac{1}{c_e^2}-\frac{1}{c_s^2}\right)\,,
\eea
where $c_e\equiv\sqrt{dP/d\rho}$ is called the equilibrium sound speed. However, Eq.~\eqref{eq:N_def} is not suitable for numerical evaluation, for the difference between $1/c_e^2$ and $1/c_s^2$ is tiny so the subtraction operation may not be accurate. Instead, we plug \cite{1994MNRAS.270..611L} Eq.~4.7 into Eq.~\eqref{eq:N_def} to get
\begin{equation}
\begin{aligned}
N^2=-\frac{\mathfrak{g}^2}{c_e^2c_s^2}\left(\frac{\partial P}{\partial Y_p}\right)_{\rho}\left(\frac{dY_p}{d\rho}\right).
\end{aligned}
\end{equation}
Because the discrepancy between $c_e^2$ and $c_s^2$ is small, it is adequate to make the following approximation
\begin{equation} \label{eq:N_calculate}
\begin{aligned}
N^2\approx -\frac{\mathfrak{g}^2}{c_e^4}\left(\frac{\partial P}{\partial Y_p}\right)_{\rho}\left(\frac{dY_p}{d\rho}\right).
\end{aligned}
\end{equation}
This equation is what we use to numerically calculate buoyancy frequency. The expression $\partial P/\partial Y_p$ at fixed $\rho$ and $dY_p/d\rho$ can be computed from the EOS.

Even though Eq.~\eqref{eq:N_def} is a universal definition for $N$, we do not use it in every part of the star. As stated in Sect.~\ref{sec:roadtoMTCS} and will be once again emphasized in Sect.~\ref{sec:staticresult}, we assume completely fluid neutron stars and confine our discussions to core g-modes. This assumption leads to the restriction that $N=0$ throughout the crust, which signifies the vanishing of crust g-mode.

What is more, under this assumptioin, it is necessary to determine the exact critical density at which the crust-core transition occurs (in other word, the position where $N$ is cut off) for each EOS.  For SLy4, since it is combined with the BBP EOS in the crust-core transition region, we adopt the critical density determined by \cite{1971NuPhA.175..225B}, with the exact value of $\rho_{\rm cut}=2.4\times 10^{14}$ $\rm g/cm^3$ (\cite{1971NuPhA.175..225B} Sect.~10). For the APR EOSs, the crust-core transition is expected to occur at $n_b = 0.1$ $\rm fm^{-3}$ (corresponds to $\rho_{\rm cut}=1.67\times 10^{14}$ $\rm g/cm^3$, see \cite{1998PhRvC..58.1804A} and \cite{1995NuPhA.584..675P} Table 1). However, for Shen, the crust-core boundary is not specified so we determine it by choosing the position where $Y_p$ hits its minimum. This is also the criterion adopted in \cite{1994MNRAS.270..611L} to distinguish the core g-modes from their crust counterparts (c.f. Sect.~4.1 in that paper).

\begin{figure}
\begin{overpic}[width=0.9\columnwidth]{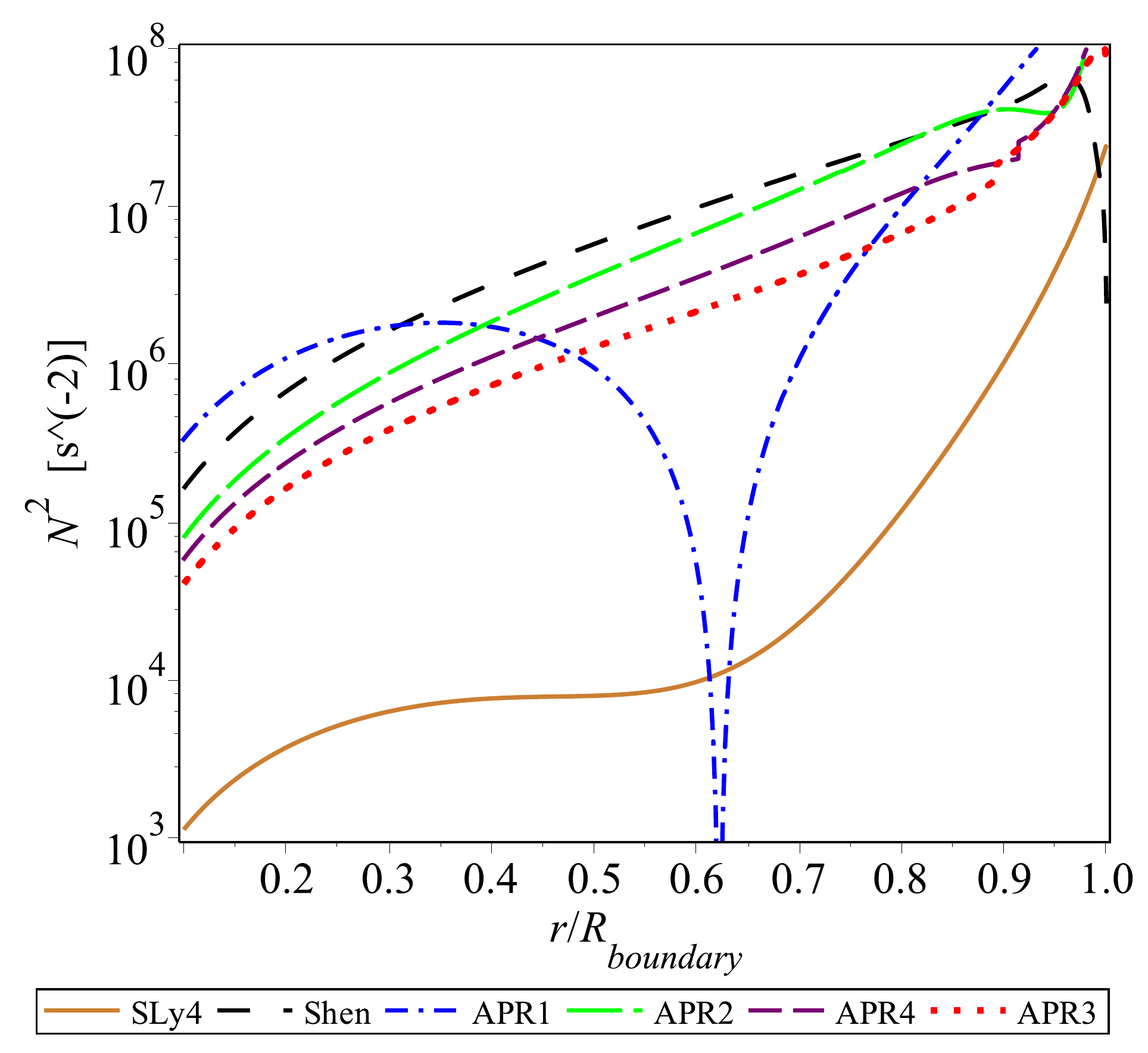}
\end{overpic}
\caption{The distribution of buoyancy frequency in the neutron star core. $R_{\rm boundary}$ is the radius where crust-core transition takes place. The discontinuities of $N^2$ in APR3 and APR4 are due to a phase transition (see Sect.~\ref{sec:APREOS}). Note that the buoyancy frequency for SLy4 model is extraordinarily small as compared to others, which has a profound effect on the MTCS that will be explained below.}
\label{fig:N^2}
\end{figure}

To conclude, we use Eq.~\eqref{eq:N_calculate} to compute $N$ in the core and set $N=0$ throughout the crust, the crust-core boundary is specified for each EOS individually. The numerical results we obtain are presented in Fig.~\ref{fig:N^2}.

\subsubsection{V}
  The external potential $V$ (in the next-to-last line of Eq.~\eqref{eq:MTCS}) is derived through the VPT. It represents, but is not exactly equivalent to, the tidal potential, and its expression is (see Eq.~48 of VZH)
\begin{equation} \label{eq:externalpoten}
\begin{aligned}
V(r)=-\frac{\omega_0^4}{10}\frac{(6-n)r^3}{\mathfrak{g}}.
\end{aligned}
\end{equation}
The quantity $n$ is defined as $d\ln \mathfrak{g}/d\ln r$, and has the value of $n\approx 1$ in the neutron star core, so we can regard it simply as a constant.

\subsubsection{$\vec{\sigma}$ and $\sigma_r$} \label{sec:radialdisplacement}

\begin{figure}
\begin{overpic}[width=0.9\columnwidth]{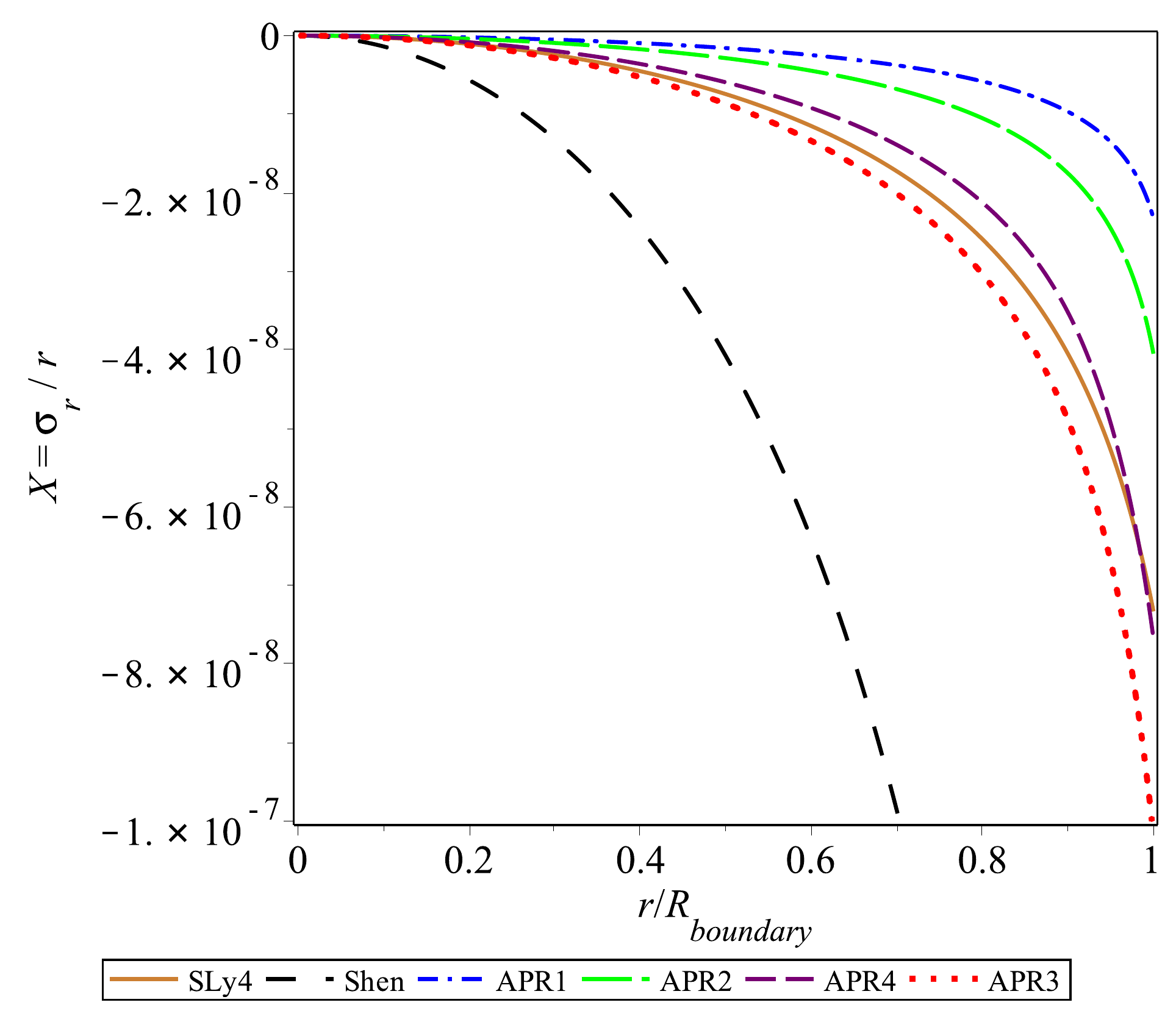}
\end{overpic}
\caption{The radial displacement measure $X=\sigma_r/r$ in the neutron star interior, as computed by solving Eq.~\eqref{eq:EOM}. Notice that the star obeying the Shen EOS has the largest displacement, which implies a comparatively severe tidal deformation. This feature was also observed in numerical simulations (see e.g.~\cite{2011MNRAS.418..427S}).}
\label{fig:rX}
\end{figure}

The vector $\vec{\sigma}$ is the displacement from unperturbed star to radially stretched spherical star (after the VPT), and $\sigma_r$ is the radial component of $\vec{\sigma}$ after separation of variables (hereafter, we will refer to it as the radial displacement). 
  
To solve the radial displacement $\sigma_r$, we use the VZH Eqs.~96 (the definition of divergence) and 98 (the radial equation of motion), which in our case are 
\begin{equation} \label{eq:radialEOM}
\begin{aligned}
(\nabla\cdot\vec{\sigma})_r=&\frac{d\sigma_r}{dr}+\frac{2}{r}\sigma_r,\\
\frac{d}{dr}[\Gamma_1P(\nabla\cdot\vec{\sigma})_r]=&-\left(\frac{2\mathfrak{g}}{r}-\frac{d\mathfrak{g}}{dr}\right)\rho\sigma_r+\rho\frac{d\Phi_{\rm tide}^\prime}{dr}\,.
\end{aligned}
\end{equation}
The validity of these equations is subtle as the radial ``mode" is not a normal mode of the star. We refer readers to the discussion in VZH for more details, but noting that only tidal perturbation to the potential $\Phi_{\rm tide}^{\prime}$ is present in Eq.~\eqref{eq:radialEOM}, and is given by the external potential in Eq.~\eqref{eq:externalpoten},
\begin{equation} \label{eq:externalV}
\begin{aligned}
\Phi_{\rm tide}^\prime=\epsilon^2V(r).
\end{aligned}
\end{equation}
The absence of $\Phi^{\prime}$ (perturbation by g-modes) is a result of the Cowling approximation\footnote{In order to be self-consistent, the computation of the radial displacement $\sigma_r$ must be confined to be under the Cowling approximation, as $\sigma_r$ is directly related to the VPT which is based on that approximation.}.
Combining the two equations in \eqref{eq:radialEOM}, we obtain
\begin{align}\label{eq:EOM}
\frac{d}{dr}&\left[\Gamma_1P_1\left(r\frac{dX}{dr}+3X\right)\right]=\left(-2\mathfrak{g}_1+r\frac{d\mathfrak{g}_1}{dr}\right)\rho X \notag \\
&-\left[\frac{(3-n)(6-n)}{10A^6}\left(\frac{GM}{c^2}\right)^2\right]\frac{\rho r^2}{\mathfrak{g}_1}\,,
\end{align}
with the dimensionless radial displacement $X\equiv \sigma_r/r$, $P_1\equiv P/c^2$, and $\mathfrak{g}_1\equiv \mathfrak{g}/c^2$. We solve this ordinary differential equation (ODE) numerically, with the neutron star EOSs listed in Sect.~\ref{sec:6EOS}, and the initial conditions of 
\bea
X(0)=0\,, \quad  X^\prime(0)=0 \,.
\eea
The condition $X(0)=0$ is a consequence of there being no radial displacements at $r=0$, while $X^{\prime}(0)=0$ is demanded by the ODE itself after setting $X(0)=0$. For demonstration, the resulting $X$ is displayed in Fig.~\ref{fig:rX}, assuming a binary separation of $A=100 \: \rm km$. We also note that to a good approximation, $X\equiv \sigma_r/r \propto A^{-6}$. Therefore, the $X$ values corresponding to other $A$ choices not shown in the figure can be estimated by a simple rescaling.

Notice that the radial displacement $\sigma_r$ is negative, indicating that the static tidal force actually compresses the star. From Fig.~\ref{fig:rX}, one notices in addition that the star governed by a soft EOS has a comparatively small radial displacement, i.e. less deformed by the tidal force. The fact that a ``soft" star is more rigid than a ``stiff" one can be explained by Figs.~\ref{fig:rrho} and \ref{fig:rP}: ``softer" stars are denser and have higher inner pressure, in other word, they {are more tightly bound.}

\subsection{The results} \label{sec:staticresult}

Now that we have discussed the quantities appearing in the MTCS, the natural subsequent step is to turn to its evaluation. However, before proceeding further, we shall rewrite Eq.~\eqref{eq:MTCS} into a different form that is more amenable to numerical evaluation. With the simplification discussed in the end of Sect.~\ref{sec:grgh}, Eq.~\eqref{eq:externalpoten} and the first equation of \eqref{eq:radialEOM}, we have that Eq.~\eqref{eq:MTCS} finally turns into
\begin{equation} \label{eq:simplifiedMTCS}
\begin{aligned}
& {\rm {MTCS}} \equiv \epsilon^2\vert2\kappa_{gg\sigma}+V_{gg}\vert \approx 
\\&
- \frac{1}{E_0} \int dr \rho g_r^2c^2 \bigg\{\bigg[\Gamma_1+1+\left(\frac{\partial\ln \Gamma_1}{\partial\ln \rho}\right)_s\bigg] \\
& \times \left(r\frac{dX}{dr}+3X\right)\frac{r^2\mathfrak{g}_1^2}{c_{s1}^2} - 4X\frac{r^2\mathfrak{g}_1^2}{c_{s1}^2}  \\
& -2r^2\left(r\Lambda_g^2\frac{\omega_g^2}{c^2}\frac{g_h^2}{g_r^2}+2\mathfrak{g}_1\right)\frac{dX}{dr} -\mathfrak{g}_1r^3\frac{d^2X}{dr^2} \\
& \times \bigg[-r^2\mathfrak{g}_1\frac{dX}{dr}-r\mathfrak{g}_1X+r^2X\frac{d\mathfrak{g}_1}{dr} \\
& -\frac{(6-n)(3-n)r^3}{10A^6\mathfrak{g}_1}\left(\frac{GM}{c^2}\right)^2\bigg]
\left(\frac{2r\mathfrak{g}_1}{c_{s1}^2}+\frac{d\ln \rho}{d\ln r}\right) \bigg\},
\end{aligned}
\end{equation}
where $c_{s1}=c_s/c$.

Now we are ready to perform the MTCS calculations for the static tide with formula \eqref{eq:simplifiedMTCS}. We impose the neutron star properties of Sect.~\ref{sec:EOS} and $g_r$, $g_h$, $X$ computed respectively in Sects.~\ref{sec:grgh} and \ref{sec:radialdisplacement}.
  The choice of binary separation is somewhat arbitrary because the MTCS, as a whole, is approximately proportional to $A^{-6}$. Furthermore, we note that the radial function $g_r$ and the dimensionless radial displacement $X$ (as well as its derivatives) are present in almost every terms in Eq.~\eqref{eq:simplifiedMTCS}, so they in fact exert significant influences on the magnitude of the MTCS. 

\begin{table}
\centering
\caption{The MTCS and the Jacobian contributions to the frequency shift, under a static tide and for $l_g=4$, $n=32$ g-modes. The neutron star radius $\mathcal{R}$ corresponding to each EOS is displayed in Tb.~\ref{tb:NSProperty}. \label{tb:staticMTCS}}
{\begin{tabular*}{0.95\columnwidth}{@{\extracolsep{\fill}}ccccc}
\toprule[2pt]
\multirow{2}{*}{EOS} & \multicolumn{2}{c}{$A=100\rm\: km$} & \multicolumn{2}{c}{$A=2\mathcal{R}$} \\

& MTCS & $\epsilon J_{gg}^{(1)}$ & MTCS & $\epsilon J_{gg}^{(1)}$ \\
\midrule[1pt]

  SLy4 & $1.30\times10^{-3}$ & $-1.08\times10^{-4}$ & 8.04 & $-8.51\times10^{-3}$ \\ 

  Shen & $1.18\times10^{-4}$ & $-2.31\times10^{-4}$ & 0.167 & $-8.70\times10^{-3}$ \\ 

  APR1 & $3.34\times10^{-5}$ & $-5.36\times10^{-5}$ & 0.857 & $-8.59\times10^{-3}$ \\ 

  APR2 & $2.20\times10^{-5}$ & $-6.05\times10^{-5}$ & 0.333 & $-7.44\times10^{-3}$ \\ 

  APR3 & $4.11\times10^{-5}$ & $-1.18\times10^{-4}$ & 0.201 & $-8.26\times10^{-3}$ \\ 

  APR4 & $2.91\times10^{-5}$ & $-9.46\times10^{-5}$ & 0.201 & $-7.85\times10^{-3}$ \\ 
\bottomrule[2pt]
\end{tabular*}}
\end{table} 

\begin{figure}
\begin{overpic}[width=0.9\columnwidth]{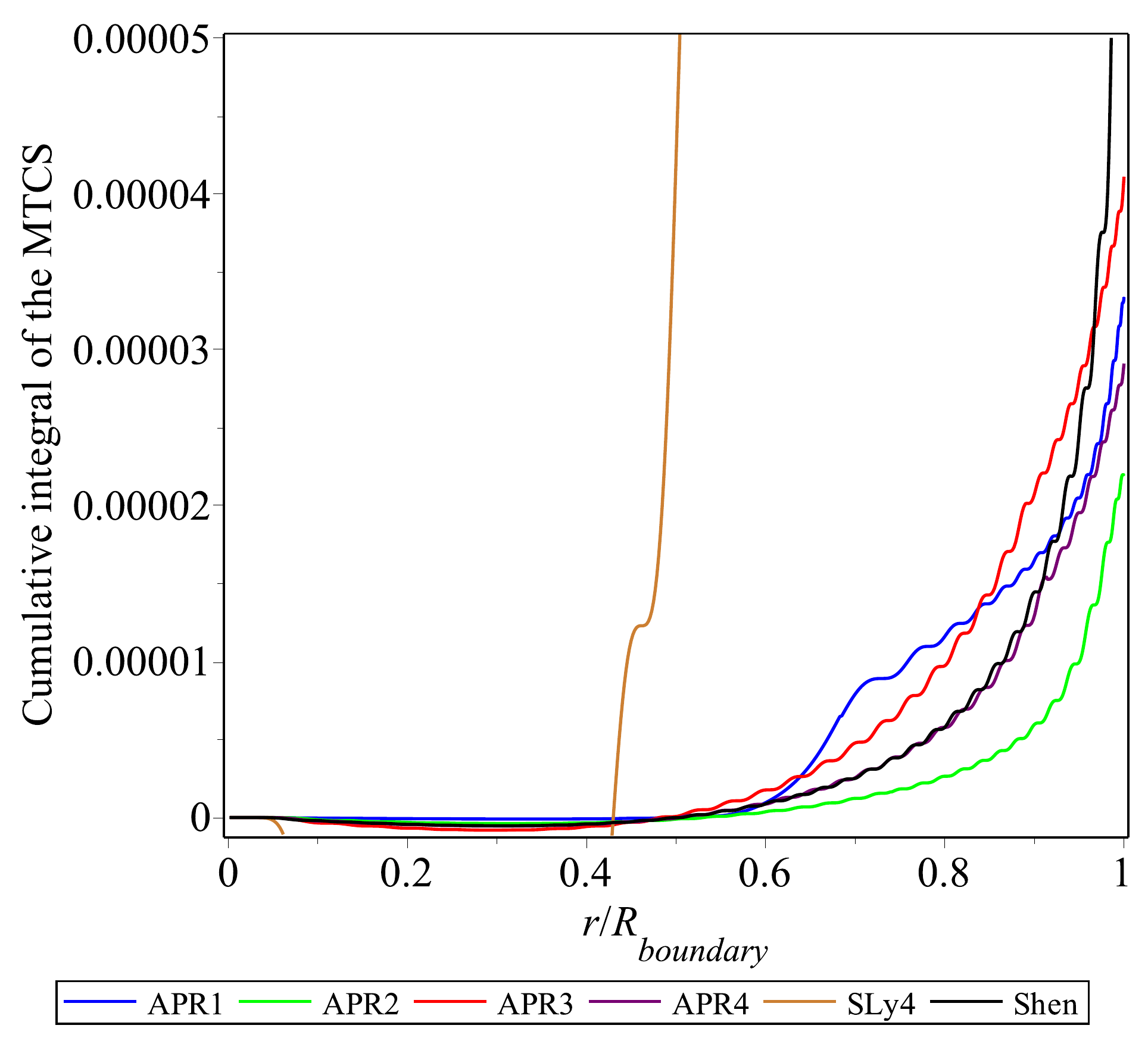}
\end{overpic}
\caption{The MTCS under static tide. The horizontal axis is the upper limit of the integration in Eq.~\eqref{eq:simplifiedMTCS}. The binary separation is set at $A=100 \:\rm km$, and all g-modes share the same degree ($l_g=4$) and radial order ($n=32$) for this figure.}
\label{fig:equiMTCS}
\end{figure}
  
  We tabulate the MTCS results for the six EOSs in Tb.~\ref{tb:staticMTCS}, and display their ``cumulative distribution'' within the neutron star in Fig.~\ref{fig:equiMTCS}, in which the horizontal axis represents the upper limit of the integral in Eq.~\eqref{eq:simplifiedMTCS} (i.e. we stop the integration prematurely at some radius before $\mathcal{R}$, to show how much different parts of the neutron star contribute to the MTCS). In WAB, it was proven that the three-mode coupling is strong in the core region (WAB Sect.~3.2). However, we see from Fig.~\ref{fig:equiMTCS}, after taking the four-mode interaction into consideration, that MTCS becomes nearly zero for all EOSs, suggesting that the cancellation between the three- and four-mode couplings as revealed by VZH is near-exact in the core. In contrast, in the outer half of the star, the near-exact cancellation begins to collapse, and MTCS grows rapidly near the crust-core interface. To explain this phenomenon, we note that a hint is provided by Fig.~\ref{fig:rX}, namely that both the (absolute) value and the slope of $X$, the frequently appearing variable in Eq.~\eqref{eq:simplifiedMTCS}, inflate significantly as $r$ approaches the crust-core interface.

Meanwhile, it should be emphasized that our treatments do not apply in the crust (a solid stratification with a density between $10^6\,\rm g/cm^3$ and the crust-core transition density $\rho_{\rm cut}$ (\cite{1983bhwd.book.....S} Ch.~9.3)) because the neutron star matter is assumed to be of a fluid nature everywhere, which is not a valid description of solid regions. Therefore, we terminate the integration in Eq.~\eqref{eq:simplifiedMTCS} at $\rho=\rho_{\rm cut}$ (geographically, $r=R_{\rm boundary}$). However, as shown by \cite{1992ApJ...395..240R}, the crust and the surface do not sustain core g-modes (c.f. Fig.~3 in that paper) so that errors induced by halting the integration before reaching $\mathcal{R}$ is unlikely to be large if we, as with the previous studies on the topic of g-mode stability, confine our discussion to core g-modes.
 
  Now that we have had a glimpse of the general characteristics of the MTCS, we turn to its EOS dependence. Since the MTCS is composed of many variables, and each of them, to a more or less extent, depends on the specific choice of the EOS, it is difficult to determine at first glance which EOS will predict a stronger MTCS and which leads to a weaker one. However, if we omit the result of APR1 and the abnormally large MTCS of SLy4 for now, and focus on the other four, we shall discover that the ranking of MTCSs for the four EOSs is in line with the ranking of their stiffness (c.f.~Sect.~\ref{sec:6EOS}). Consequently, one may speculate that the coupling to tide tends to be stronger in stars with stiffer EOSs. Indeed, as discussed in Sect.~\ref{sec:radialdisplacement} and depicted in Fig.~\ref{fig:rX}, stars predicted by softer EOSs are less severely deformed by the tidal force, i.e., have smaller $X$. To be more specific, we can use APR2 (a soft EOS) and APR3 (a relatively stiff EOS) for comparison (from the same family, thus cleaner comparison): Our numerical result shows that in most parts of the core, the dimensionless radial displacement (as well as its first and second derivatives) for stars governed by APR3 is roughly twice that given by the APR2 EOS, which is consistent with the value of MTCS for the APR3 model being approximately twice greater than its APR2 counterpart.
  
  Besides greater $X$, a stiff EOS also leads to larger inhomogeneous terms that contain the external potential $V$. Specifically, the inhomogeneous term in Eq.~\eqref{eq:simplifiedMTCS} is
\begin{equation}
\begin{aligned}
-\frac{(6-n)(3-n)}{10A^6}\left(\frac{GM}{c^2}\right)^2\frac{r^3}{\mathfrak{g}_1}\propto\mathfrak{g}_1^{-1}.
\end{aligned}
\end{equation}
Applying the TOV equations \eqref{eq:TOV}, we further obtain that $\mathfrak{g}_1 \propto M_{\odot}m_{*}+4\pi r^3P_1$.
Since the star mass is fixed in our computation, the equation above then tells us that a softer EOS with a higher interior pressure (see Fig.~\ref{fig:rP}) corresponds to a smaller inhomogeneous term (as expected, because the gravitational acceleration is stronger in more compact stars predicted by softer EOSs).

So far, the picture is that stiffer EOSs predict less compact and larger neutron stars (Fig.~\ref{fig:rrho} and Tb.~\ref{tb:NSProperty}), which are more easily deformed by tidal forces (Fig.~\ref{fig:rX}) and possess larger inhomogeneous terms, leading to stronger mode-tide couplings. However, our analysis is not complete yet, as we have deliberately overlooked the results for APR1 and SLy4. For these two EOSs, there is no apparent correspondence between stiffness and the MTCS. Especially striking is the MTCS for SLy4; regarding stiffness, SLy4 is only an intermediate EOS, yet it predicts extremely strong coupling that is far in excess of the other five. Hence, there must exist other (at times more dominant) factors to account for the conspicuously large MTCS for SLy4. 

A careful scrutinization of Eq.~\eqref{eq:simplifiedMTCS} reveals that $g_r$, the radial component of the g-mode eigenfunction, also plays an important role in determining the mode-tide coupling. This is physically reasonable as larger intrinsic modal deformations should enable greater overlaps with tidal deformations. The question that arise then is: when the degree ($l_g$) and the radial order ($n$) of the g-mode is fixed, what kind of EOSs will give a larger g-mode amplitude? To answer this question, we employ the Wentzel-Kramers-Brillouin (WKB) method to provide analytic forms for the g-mode eigenfunction. The WKB method (for a detailed introduction to the method and its applications to stellar oscillations, please refer to standard textbooks such as \cite{CDBook}) is a satisfactory approximation for high frequency p-modes and low frequency g-modes (we will only use it for explanatory illustrations here, so the results in this paper are not bound by the applicability of the WKB).  Under the Cowling and the WKB approximations, the solution of Eq.~\eqref{eq:governingequation_gr} is
\begin{equation} \label{eq:WKB}
\begin{aligned}
&g_r\simeq \frac{A_g}{N}\sin(k_gr)=\sqrt{\frac{E_0\alpha_g}{\rho r^2}}\frac{\sin(k_gr)}{N},\\
&g_h\simeq \frac{A_g}{\omega_g\Lambda_g}\cos(k_gr)=\sqrt{\frac{E_0\alpha_g}{\rho r^2}}\frac{\cos(k_gr)}{\omega_g\Lambda_g},
\end{aligned}
\end{equation}
where (WAB Sect.~3.2)
\bea \label{eq:alpha_g}
\quad \alpha_g\equiv \frac{N}{r}\left(\int Nd\ln r\right)^{-1}\approx \frac{0.4}{\mathcal{R}}\,, \quad
\eea
and $k_g$ is the g-mode wave number given by $k_g\simeq \Lambda_gN/(r\omega_g)$. From Eqs.~\eqref{eq:WKB} and \eqref{eq:alpha_g}, we have that the factor $\rho g_r^2/E_0$ in Eq.~\eqref{eq:simplifiedMTCS}, as a whole, is roughly in inverse proportion to $N^2$. That is, when other variables are controlled, EOSs with smaller buoyancy frequency through the neutron star core are expected to sustain larger g-mode amplitudes, resulting in stronger couplings. We are now able to provide an explanation for the case of SLy4. Fig.~\ref{fig:N^2} shows that the $N^2$ calculated with SLy4 is two orders of magnitude smaller than with the other EOSs across nearly the entire core region. Accordingly, the inverse proportion relation then gives much greater $\rho g_r^2/E_0$ as compared to the other models, and subsequently a large MTCS.

  To sum up the relationship between static MTCSs and the EOSs: two dominant ingredients, the stiffness of the EOS and the buoyancy frequency predicted by the EOS, affect the MTCS simultaneously. In other words, the mode-tide coupling depends on the description of high density nuclear matter in a rather sophisticated way, both the property of nuclear matter in $\beta$-equilibrium (i.e., the stiffness of EOS) and away from equilibrium (represented by the buoyancy frequency) would have an impact on the coupling strength.

  Now that the properties of the MTCS have been discussed, we briefly turn to the Jacobian term $\epsilon J_{gg}^{(1)}$. According to Tb.~\ref{tb:staticMTCS}, at $A=100\rm\: km$, $\epsilon J_{gg}^{(1)}$ is approximately one tenth of the MTCS in the SLy4 model, comparable with the MTCS for Shen, while several times the size of the MTCS for the APR EOSs. However, since $\epsilon J_{gg}^{(1)}\propto A^{-3}$ (see Eq.~\eqref{eq:Jgg}) while ${\rm MTCS}\propto A^{-6}$, the MTCS gradually gain dominance as the binary winds tighter.
As for the EOS dependence, like MTCS, the exact values of $\epsilon J_{gg}^{(1)}$ also vary from one EOS to another: the stiffer the EOS we choose, the larger the $\epsilon J_{gg}^{(1)}$ we get.

  Aside from examining the MTCS's (as well as the Jacobian's) dependence on the EOSs, we also confirm VZH's result directly: that the non-resonant instability does not occur in the case of static tide during early inspiral stage (VZH Sect.~4.2). Over and above that, VZH's assertion can be extended to the entire evolution process of neutron star binaries. To achieve this, we compute the MTCSs as well as the Jacobians in the extreme situation of $A=2\mathcal{R}$, when the two stars are well into the merger phase. Nevertheless, from Tb.~\ref{tb:staticMTCS}, we see that both $\rm MTCS$ and $\epsilon J_{gg}^{(1)}$ are less than $1$ for Shen and APR1-4 EOSs (with SLy4 the only exception), and
\begin{equation} \label{eq:peromega1}
\begin{aligned}
\frac{\omega_-^2}{\omega_g^2}\approx 1-\left\vert2\epsilon J_{gg}^{(1)}+\epsilon^2\vert2\kappa_{gg\sigma}+V_{gg}\vert\right\vert>0\,,
\end{aligned}
\end{equation}
representing stable g-modes. In other words, for most EOSs, the onset of instability {due to the static tide is avoided} all the way up to merger, not only during the early inspiral period when two stars are far apart. It should of course be pointed out here that our formalism actually breaks down at $A=2\mathcal{R}$. At such a close distance, higher order tidal potentials should be included. For example, the $l=3$ (octupole) term is no longer negligible. However, higher order terms are always smaller than the leading order (in the case of $A=2\mathcal{R}$, the octupole term is less than half of the quadrupole one). We therefore expect the qualitative, although not the quantitative, conclusion to remain valid even with a more thorough treatment of the merger phase. 
  
In this section, we have demonstrated that nonlinear mode couplings to the {static tide} do depend on neutron star EOS, in fact rather sensitively. However, the baseline MTCS value is very low, so the nonlinear effect is likely too small to be detectable in the static tide scenario. Interestingly though, more recent results by W2016 suggest that the situation appears to be different when one considers the non-static tide (through the inclusion of the $m=\pm 2$ harmonics). We turn to this case below, and show that the sensitive EOS dependence is preserved (as is to be expected because the qualitative consequences of the differences in EOSs invoked so far do not rely on the tide being static).
  
\section{Non-static Tide} \label{sec:Dynamic}

  {We turn now to the non-static tide, and show that the temporal variation introduces highly nontrivial effects, and so the behaviour of the resulting MTCS changes significantly. Above all, we emphasize the compressible nature of the non-static tide in Sect.~\ref{sec:intronst}, demonstrating that it does not preserve the volume of the star at leading order and may therefore already result in the resurrection of the instability at this order. We then move on to evaluate the first order MTCS in Sect.~\ref{sec:fsft_new} to verify that this is indeed the case. It turns out that it is not straight-forward, at least not in an airtight rigorous manner, to adapt the VPT method to the time dependent case, so we evaluate the MTCS using its non-VPT-treated original expression instead. With the results thus obtained, we discuss in Sect.~\ref{sec:nst_result} the general features of the MTCS and the sensitive EOS dependence it exhibits, before interpreting the observational consequences of these findings in Sect.~\ref{sec:instability}.}

\subsection{The compressible nature of non-static tide} \label{sec:intronst}
     
  Non-static tides are more realistic for inspiraling binaries. In this context, the tidal force comes from a companion star in circular motion rather than standing still. The spherical harmonic expansion of the full tidal potential $\epsilon U_{\rm full}$ is (\cite{1994MNRAS.270..611L} Eq.~2.2)
\begin{equation} \label{eq:dyntideori}
\begin{aligned}
\epsilon U_{\rm full}=-GM\sum \limits_{l,m}W_{lm}\frac{r^l}{A^{l+1}}Y_{lm}(\theta,\phi)e^{-im\Omega t}\,.
\end{aligned}
\end{equation}
Same as W2016, we keep the leading quadrupole term (W2016 Eq.~3) to get 
\begin{equation} \label{eq:dyntide}
\begin{aligned}
U_{\rm full}\approx -\omega_0^2r^2\sum\limits_{m=-2}^2W_{2m}Y_{2m}(\theta,\phi)e^{-im\Omega t},
\end{aligned}
\end{equation}
where the coefficients $W_{lm}$ depend on $(l,m)$, with in particular $W_{20}=-\sqrt{\pi/5}$, $W_{2,\pm 1}=0$, and $W_{2,\pm 2}=\sqrt{3\pi/10}$. The $m=0$ term in Eq.~\eqref{eq:dyntide} is the previously discussed static tide while the $m=\pm 2$ harmonics refer to the non-static tide, as characterized by the presence of the orbital angular frequency $\Omega$.
Although the differences between static and non-static tides come only through the $e^{-im\Omega t}$ factor, the latter is nevertheless accompanied by important new effects. For example, the non-static tide is now able to (a) excite g-modes in compact binaries by the resonant excitation mechanism \citep{1994MNRAS.270..611L,2011MNRAS.412.1331F}; (b) change the volume of the star at leading order (W2016). Although observation (a) has an impact on the nonlinear mode coupling (see W2016 Fig.~9, it accounts for the fluctuation in the MTCS with binary separation), it is (b) that spoils the near-exact cancellation and raises the MTCS orders of magnitude larger. Here we go into some details and provide a brief description for the consequences of (b).

  Beginning with the basic hydrodynamic equations and the non-static tidal potential, W2016 gave the expression for the first-order (in $\epsilon$, similarly hereinafter) tidal displacement $\vec{\chi}^{(1)}$ and showed that (W2016 Eq.~62)
\begin{equation} \label{eq:divchi}
\begin{aligned}
\nabla\cdot \vec{\chi}^{(1)}\simeq \frac{1}{2}\left(\frac{m\Omega}{N}\right)^2\frac{\mathfrak{g}}{c_s^2}\frac{d\ln \rho}{d\ln r}\chi_{r,\rm st}^{(1)}\,,
\end{aligned}
\end{equation}
where $\chi_{r,\rm st}^{(1)}$ is the radial component of the first-order tidal displacement we had with static tides. With Eq.~\eqref{eq:divchi}, it's obvious that $\nabla\cdot \vec{\chi}^{(1)}=0$ in a static tide ($\Omega=0$), and so we had an invariant volume at leading order. In other words, after VPT, the new spherical star is exactly the same as the unperturbed one (at leading order), meaning there is no first-order radial displacement (i.e., $\sigma_r^{(1)}=0$). Consequently, the first-order external potential $V^{(1)}$ vanishes (see Eq.~\eqref{eq:externalV}, the tidal potential enters at the second-order). On the other hand, $\Omega\neq 0$ leads to a nonvanishing divergence for $\vec{\chi}^{(1)}$  (termed the ``finite frequency correction to linear tide" in W2016) and subsequently nonvanishing $\sigma_r^{(1)}$ and $V^{(1)}$ (elaborated in Appendix \ref{app:VPT}) that alters the situation significantly. Namely, as noted in Sect.~\ref{sec:staticresult}, the radial displacement and the inhomogeneous term are two crucial constituents in the expression of MTCS and make non-negligible contributions to the coupling strength, so one should thus expect the MTCS to become roughly $1/\epsilon$ times larger in a non-static setting.

To investigate the EOS dependence, explicit computations of the MTCS are needed. In principle, both the first and second order contributions to the frequency shift in Eq.~\eqref{eq:perturbedgmode1} should be evaluated.
The second order frequency shift which contains the four-mode coupling constant and the second order tidal displacements had already been carefully scrutinized by W2016, and its effects are embodied by the three- and four-mode residual term $R_{gg}$ in that paper. We note further though, that since $\nabla \cdot \vec{\chi}^{(1)} \neq 0$, $\sigma_r^{(1)} \neq 0$ and $V^{(1)} \neq 0$, the first order contribution also ceases to be an insignificant piece in the case of non-static tide.  
To {quantify} this leading order effect (not always greater in value than higher order terms though, as we will elaborate below), we {compute the first order MTCS using the formalism developed by \cite{2012ApJ...751..136W} (WAQB). This more direct but also more computationally intensive method is needed due to the difficulties in generalizing the VPT to the time dependent case (see Appendix \ref{app:VPT} for details).}   

\subsection{{The first order mode-tide coupling strengths}} \label{sec:fsft_new}

\begin{figure}
\begin{overpic}[width=0.95\columnwidth]{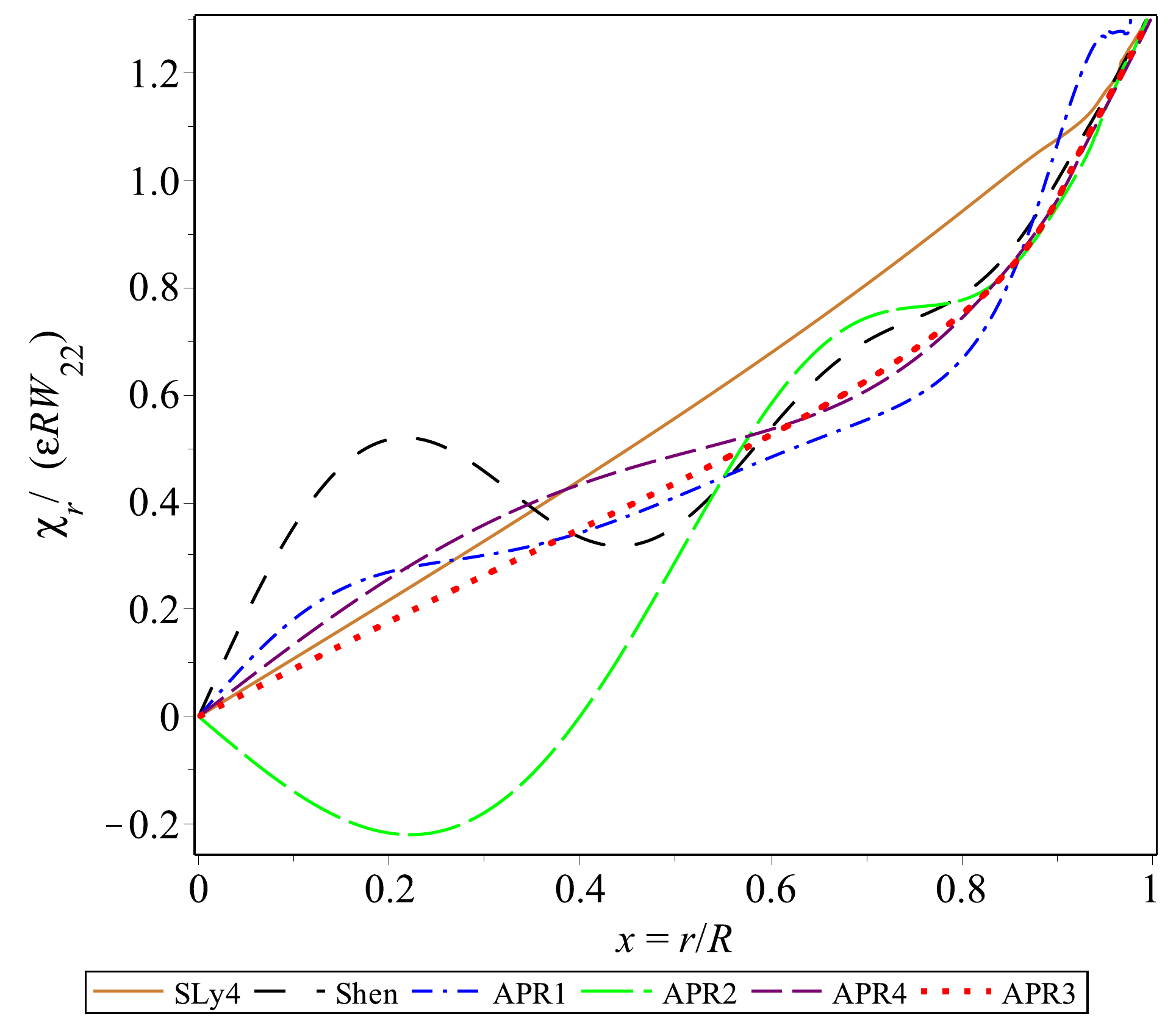}
\end{overpic}
\caption{The (scaled) linear tidal displacement computed under the $m=2$ non-static tide at binary separation $A=150 \; \rm km$. Since $M=1.4\rm M_{\odot}$ is fixed for all six EOSs, this distance corresponds to the GW frequency $f_{\rm gw}=(1/\pi)\sqrt{2GM/A^3}\approx 106\rm Hz$.}
\label{fig:chi_r}
\end{figure}

{Because a rotating tidal force does not preserve the volume of a star even at the linear order in $\epsilon$, we neglect all higher order contributions in the rest of this chapter, and concentrate on expounding how the first order MTCS is computed.}
   
 {To begin with, we note that the g-mode frequency shift resulting from tidal perturbations, keep to leading order, is given by (Eq.~\eqref{eq:perturbedgmode1})
\begin{equation}
\frac{\omega_-^2}{\omega_g^2} = 1-\epsilon \left\vert U_{gg} + 2\kappa_{\chi^{(1)}gg}\right\vert + \mathcal{O}(\epsilon^2).
\end{equation}
Just as in Sect.~\ref{sec:roadtoMTCS}, the absolute value of the first order MTCS (hereinafter simply referred as the MTCS) is taken because we are only concerned with the frequency shift that may lead to instabilities. Meanwhile, as mentioned in WAQB and VZH, the \textit{g-mode, g-mode, tide} coupling constant $\kappa_{\chi^{(1)}gg}$ can be separated into a homogeneous part in which the tidal displacement is treated using normal modes, and an inhomogeneous part that accounts for the tidal perturbation to the gravitational potential, specifically
\begin{equation}
\frac{\omega_-^2}{\omega_g^2} = 1-\epsilon \left\vert U_{gg} + 2\kappa_{\chi^{(1)}gg, I} + 2\kappa_{\chi^{(1)}gg, H}\right\vert + \mathcal{O}(\epsilon^2).
\end{equation}
The homogeneous three-mode coupling constant $\kappa_{\chi^{(1)}gg,H}$ was originally derived in WAQB Appendix A while the rest of the MTCS is given by WAQB Eq.~A71. Here we rearrange and simplify them under the Cowling approximation, into
\bea
&& {\rm MTCS} \equiv \epsilon \vert U_{gg} + 2\kappa_{\chi^{(1)}gg, I} + 2\kappa_{\chi^{(1)}gg, H} \vert \approx 
\nonumber \\
&& \epsilon \left\vert \frac{1}{E_0}  \int dr \left\lbrace
T \rho r^2 c_s^2 \left[ \Gamma_1 + 1 + \left( \frac{\partial \ln\Gamma_1}{\partial \ln \rho} \right)_s \right]
\nabla \cdot \vec{\chi}^{(1)} (\nabla \cdot \vec{g})_r^2
\right.\right. \nonumber \\ && \left.\left.
+ T \rho r c_s^2 \left[ 2\nabla \cdot \vec{\chi}^{(1)} (\nabla \cdot \vec{g})_r \left( g_h \Lambda_g^2 - 4g_r \right)
\right.\right.\right.  \nonumber \\ && \left.\left.\left.
+ (\nabla \cdot \vec{g})_r^2 \left( \chi_h^{(1)} \Lambda_{\chi}^2-4\chi_r^{(1)}  \right) \right]
\right.\right. \nonumber \\ && \left.\left.
+ T \rho \frac{d\ln\rho}{d\ln r} \left( 4\mathfrak{g} + r\frac{d\mathfrak{g}}{dr} \right)  \chi_r^{(1)} g_r^2
\right.\right. \nonumber \\ && \left.\left.
+ T \rho r \left( 4\mathfrak{g} + r\frac{d\mathfrak{g}}{dr} \right) \left[ \nabla \cdot \vec{\chi}^{(1)}  g_r^2 + 2 (\nabla \cdot \vec{g})_r g_r\chi_r^{(1)} \right]
\label{eq:nstMTCS_l4}
\right.\right. \\ && \left.\left.
-\rho r \chi_h^{(1)} g_h g_h \left( \omega_{\chi}^2 G_{\chi} + \omega_g^2 G_g + \omega_g^2 G_g \right)
\right.\right. \nonumber \\ && \left.\left.
- \rho r \chi_r^{(1)} g_h g_h \left[  (\omega_{\chi}^2-3\omega_g^2-3\omega_g^2)F_{\chi} - 2(\omega_g^2F_g+\omega_g^2 F_g) \right]
\right.\right. \nonumber \\ && \left.\left.
- \rho r \chi_h^{(1)} g_r g_h \left[  (\omega_g^2-3\omega_g^2-3\omega_{\chi}^2)F_g - 2(\omega_g^2F_g+\omega_{\chi}^2 F_{\chi}) \right]
\right.\right. \nonumber \\ && \left.\left.
- \rho r \chi_h^{(1)} g_r g_h \left[ (\omega_g^2-3\omega_{\chi}^2-3\omega_g^2)F_g - 2(\omega_{\chi}^2F_{\chi}+\omega_g^2 F_g) \right]
\right.\right.  \nonumber \\ && \left.\left.
+ \rho r \chi_h^{(1)} g_r g_r \left( \omega_g^2 F_g + \omega_g^2 F_g - 6\omega_{\chi}^2 T \right)
\right.\right. \nonumber \\ && \left.\left.
+ \rho r \chi_r^{(1)} g_r g_h \left( \omega_g^2 F_g + \omega_{\chi}^2 F_{\chi} - 6 \omega_g^2 T \right)
\right.\right. \nonumber \\ && \left.\left.
+ \rho r \chi_r^{(1)} g_r g_h \left( \omega_{\chi}^2 F_{\chi} + \omega_g^2 F_g - 6 \omega_g^2  T \right)
\label{eq:nstMTCS_l7}
\right\rbrace 
\right. \\ && \left.
- W_{lm} \frac{T(l+2)}{M\mathcal{R}^l}\int dr \rho r^l \left[ \frac{\partial \ln \rho}{\partial \ln r}g_r^2 + 2r g_r (\nabla \cdot\vec{g})_r \right] \right\vert,
\label{eq:Jplus2kapI}
\eea
where $\omega_{\chi} \equiv m\Omega$ and the subscript ``$\chi$'' denotes those entities relevant to the tide. The quantities $T$, $F_a$ and $G_a$ on the other hand, are angular integrals defined via Eq.~\eqref{eq:angularint}. The expression above is applicable to both static and non-static tides. In the former case, we have zero orbital frequency ($\Omega = 0$) and $\nabla \cdot \vec{\chi}^{(1)}=0$, while $\chi_r^{(1)}$, $\chi_h^{(1)}$ are given analytically in Eq.~\eqref{eq:chirh}. Using these constrains to further simplify Eqs.~\eqref{eq:nstMTCS_l4}-\eqref{eq:Jplus2kapI}, one will find that lines \eqref{eq:nstMTCS_l4} and \eqref{eq:Jplus2kapI} cancel out and the MTCS reduces into lines \eqref{eq:nstMTCS_l7} (with $\omega_{\chi}=0$) which is comparable to the first order Jacobian $\epsilon J_{gg}^{(1)}$ in magnitude. Therefore, under static tide, no instability will occur at leading order (nor in second order, as concluded in Sect.~\ref{sec:Static}). In contrast, under the circumstance of a non-static tide, although the definitions of $\vec{\chi}$ and $\vec{g}$ remain similar to the static case, they pick up an additional exponential factor $e^{-im\Omega t}$. Carrying out the separation of variables in the same way as in W2016 Eq.~50, we obtain
\begin{equation}
\begin{aligned}
\vec{\chi}&\equiv \left[\chi_r(r)\hat{r}+r\chi_h(r)\nabla_h\right] Y_{lm}e^{-im\Omega t}\,,\\
\vec{g}&\equiv \left[g_r(r)\hat{r}+rg_h(r)\nabla_h\right] Y_{lm}e^{-im\Omega t}\,.\\
\end{aligned}
\end{equation}
This innocuous-looking alteration to $\vec{\chi}$ induce deep-reaching changes in the magnitude of MTCS. Specifically, it spoils the exact cancellation between lines \eqref{eq:nstMTCS_l4} and line \eqref{eq:Jplus2kapI}, thereby raises the MTCS orders of magnitudes larger. The finite orbital frequency also makes analytic solutions for the linear tide difficult to acquire, so instead, the radial and horizontal components of the linear tide, i.e.~$\chi_r^{(1)}$ and $\chi_h^{(1)}$, are now obtained numerically by solving the forced oscillation equation (similar to the free oscillation equations \eqref{eq:governingequation_gr}, but contain tidal forcing terms, c.f.~Eq.~\eqref{eq:forcedODE} in Appendix \ref{app:numchi}) at a fixed binary separation $A$, using the shooting technique (also introduced in Appendix \ref{app:numchi}). The result for $\chi_r^{(1)}$ at $A = 150 \; \rm km$ is plotted in Fig.~\ref{fig:chi_r}. Other ingredients in the MTCS, including neutron star properties, g-mode eigenfrequencies and eigenfunctions, are also evaluated utilizing numerical approaches (c.f.~Sect.~\ref{sec:ingredMTCS}). All these ingredients are then substituted into Eqs.~\eqref{eq:nstMTCS_l4}-\eqref{eq:Jplus2kapI} to yield the MTCS under non-static tides.
}

\subsection{{Results and discussions}} \label{sec:nst_result}

\begin{figure*}
\subfigure{  \begin{overpic}[width=0.95\textwidth]{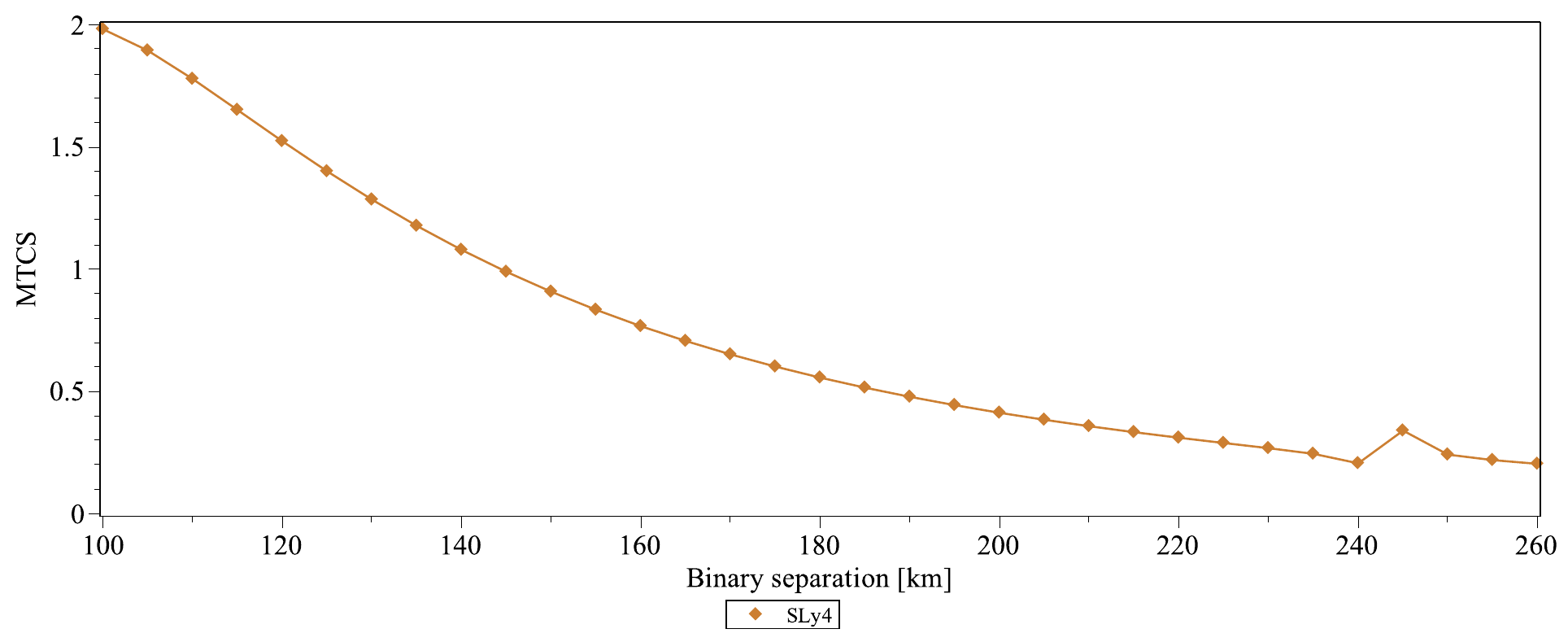}
\end{overpic}  }
\subfigure{  \begin{overpic}[width=0.95\textwidth]{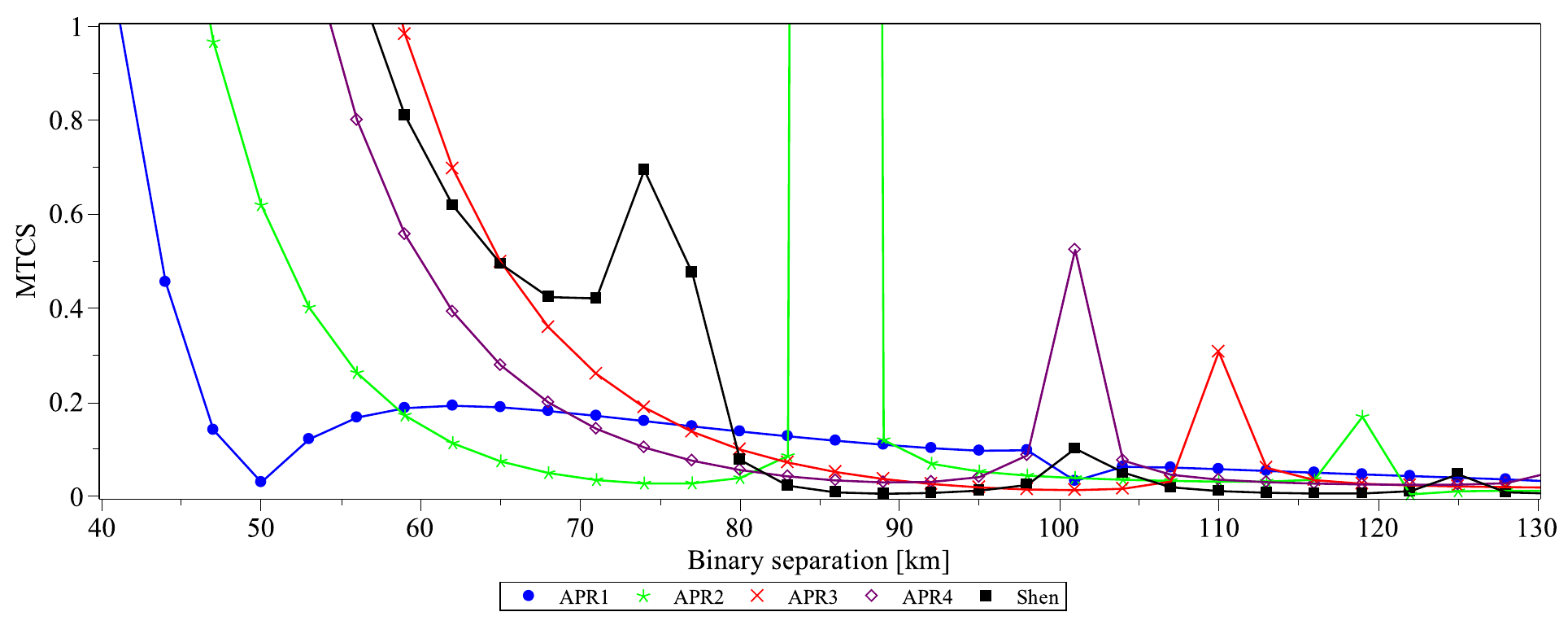}
\end{overpic}  }
\caption{The MTCSs computed for six EOSs under a non-static tide. SLy4 is depicted in the top panel while the others in the bottom (separated due to differences in scale). Horizontal axes are binary separations and vertical axes are the corresponding MTCS values. Notice that for this plot, the example g-modes have degree $l_g=4$ and radial order $n=32$. Also worth mentioning are the peaks shown in this figure, which originate from the resonance between the driving frequency (of the tide) and the intrinsic vibration frequencies (the normal modes) of the star.}
\label{fig:nstMTCS}
\end{figure*}

\begin{figure*}
\begin{overpic}[width=0.95\textwidth]{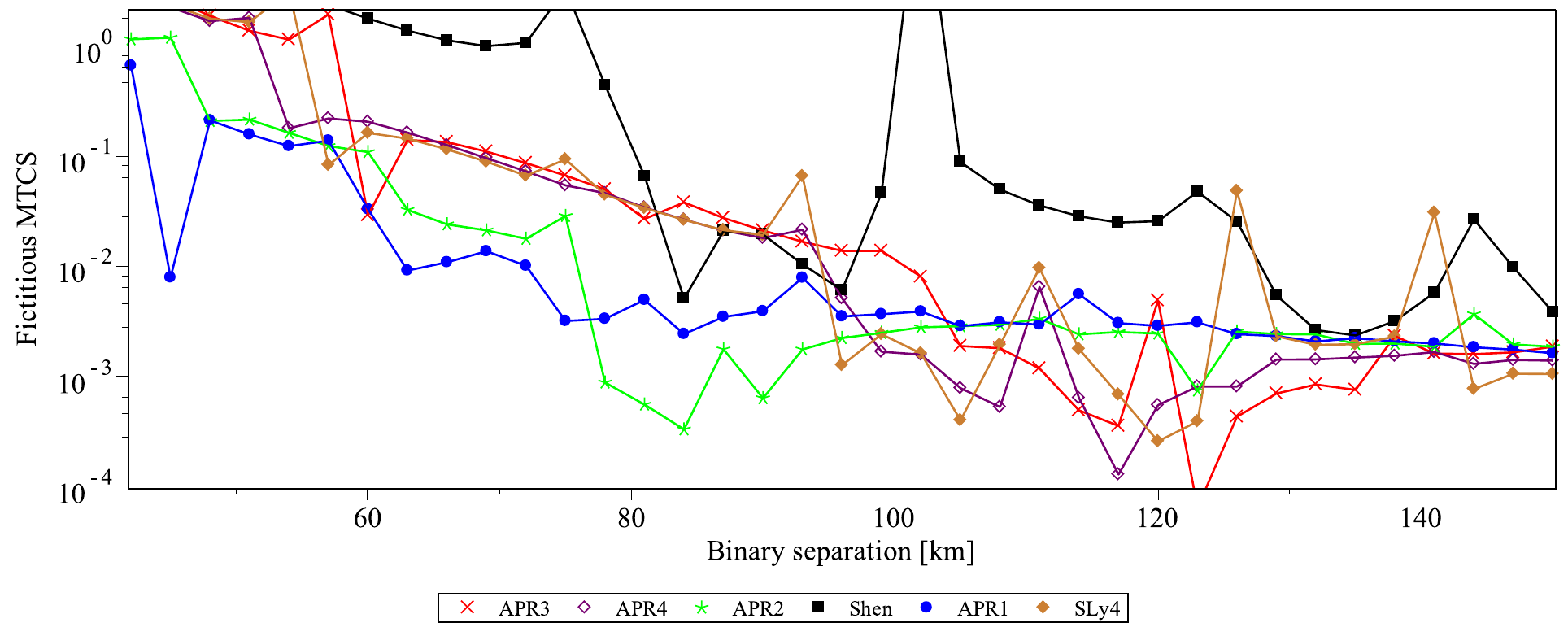}
\end{overpic}
\caption{The MTCSs calculated by controlling all six EOSs to possess exactly the same buoyancy frequency $N$ throughout the star. Note the MTCS values appearing in this figure are fictitious, constructed to isolate the effects of the EOSs stiffness, and will not arise in real astronomical settings.}
\label{fig:hypoMTCS}
\end{figure*}

The coupling strengths, as a function of binary separation, are illustrated in Fig.~\ref{fig:nstMTCS}. The mode-tide coupling becomes stronger when the binaries orbit closer. Nevertheless, at certain distances, the MTCS soars to extraordinarily large values (see those peaks in Fig.~\ref{fig:nstMTCS}). These sudden changes in the coupling strength are caused by the resonance between the non-static tide and the normal modes of the neutron star. Such resonances were also observed in W2016 (c.f.~Fig.~1 in that paper) and have been discussed in \cite{1994MNRAS.270..611L} in detail. Moreover, as mentioned in Sect.~\ref{sec:intronst}, our computations are in the leading order in $\epsilon$, while the second order term $R_{gg}$ was investigate by W2016. What they found is an inverse square relation between $R_{gg}$ and the g-mode angular frequency $\omega_g$ (W2016 Eq.~111)
\begin{equation}
\begin{aligned}
\vert R_{gg} \vert \simeq \lambda^2 \epsilon^2 \frac{\omega_0^2}{\omega_g^2},
\end{aligned}
\end{equation}  
where $\lambda$ is a parameter varying roughly between $0.1$ and $10$ (see W2016 Fig.~9), evaluated by fitting the expression above to numerical results. Setting the g-mode eigenfrequency to $f_g = 2.71 \: \rm Hz$, the binary separation to $A = 12 \mathcal{R}$ and the SLy4 radius to $\mathcal{R} = 11.663\rm \: km$, and also noting that the corresponding $\lambda^2$ from W2016 Fig.~8 is 0.2, we have that the W2016 result for $R_{gg}$ is $\approx 0.027$ for the $n=32$, $l_g=4$ g-mode choice. This value is about one-fortieth of the corresponding first order term from our MTCS computation, which turns out to be {$1.08$} at the same binary separation. Therefore, for moderately high order p-g pairs, g-mode frequency shift is dominated by the first order term. In contrast, the magnitude hierarchy between the MTCS and $R_{gg}$ flips when we consider very high order g-modes with angular frequencies $\omega_g \lesssim 1 \: \rm rad/s$, because $R_{gg}$ scales as $\omega_g^{-2}$ while MTCS is not sensitive to $\omega_g$ (one can recognize this by substituting Eq.~\eqref{eq:WKB} into Eq.~\eqref{eq:nstMTCS_l4}-\eqref{eq:Jplus2kapI}).

  More central to our study, we find that the EOS-dependence of the MTCS remains strong in the context of non-static tides: Given the observation in Fig.~\ref{fig:nstMTCS}, the MTCS for the SLy4 model surpasses the others by decades. What is more, the character of this dependence is also the same as before: the stiffness of the EOS and the buoyancy frequency predicted by the EOS affect the MTCS simultaneously. By comparing the two panels in Fig.~\ref{fig:nstMTCS}, it is plain to see that EOSs with smaller buoyancy frequencies, such as the SLy4, yield more intense mode-tide couplings. However, the dependence of the MTCS on the stiffness of the EOSs is not as straightforward to {perceive} from this figure. To make manifest the influence of the stiffness on the MTCS, we control the variable $N$, i.e., calibrate the buoyancy frequencies in different EOSs to share the same values throughout the star, and display the corresponding fictitious MTCSs in Fig.~\ref{fig:hypoMTCS}. With the aid of Fig.~\ref{fig:hypoMTCS}, we notice that the coupling constant depends on the stiffness of the EOSs only moderately. For instance, the MTCS predicted by the stiffest Shen EOS is a few times larger (therefore overwhelmed by the buoyancy frequency-effect, which can differ by orders of magnitudes between EOSs) than that given by the soft APR1 and APR2. Moreover, as revealed by numerical computations, the g-mode eigenfunctions $g_r$ and $g_h$ are not very sensitive to stiffness. Rather, the stiffness of the EOS affect the mode-tide coupling mainly through its influence on tidal deformation. This feature has also seen during previous investigations. For example, with dynamical tide, \cite{2012PhRvD..86d4032M} discovered that the tidal Love number $k_2$\footnote{The tidal Love number $k_2$ is defined via $k_2=-(3Q_{ij})/(2\mathcal{R}^5C_{ij})$, where $Q_{ij}$ is the quadrupole moment tensor of the star and $C_{ij}$ is the tidal field tensor.}, which quantitatively measures the extent of deformation of a star due to the external tidal field, is larger for a stiffer EOS during late inspiral (c.f.~Fig.~1 in \cite{2012PhRvD..86d4032M}). Similar conclusions were also made when a static tidal field is assumed (consult, e.g., the pioneer work by \cite{2010PhRvD..81l3016H} regarding this issue).
  
  In short, the MTCS turns out to be quite sensitive to the neutron star EOS, namely that MTCSs predicted by different EOSs can vary by over a decade. This observation, and the related possible instability (detailed below), offer an intriguing opportunity to distinguish EOSs by examining the nonlinear couplings' effects on binary coalescences.

\subsection{The instability} \label{sec:instability}

\begin{figure}
\begin{overpic}[width=0.9\columnwidth]{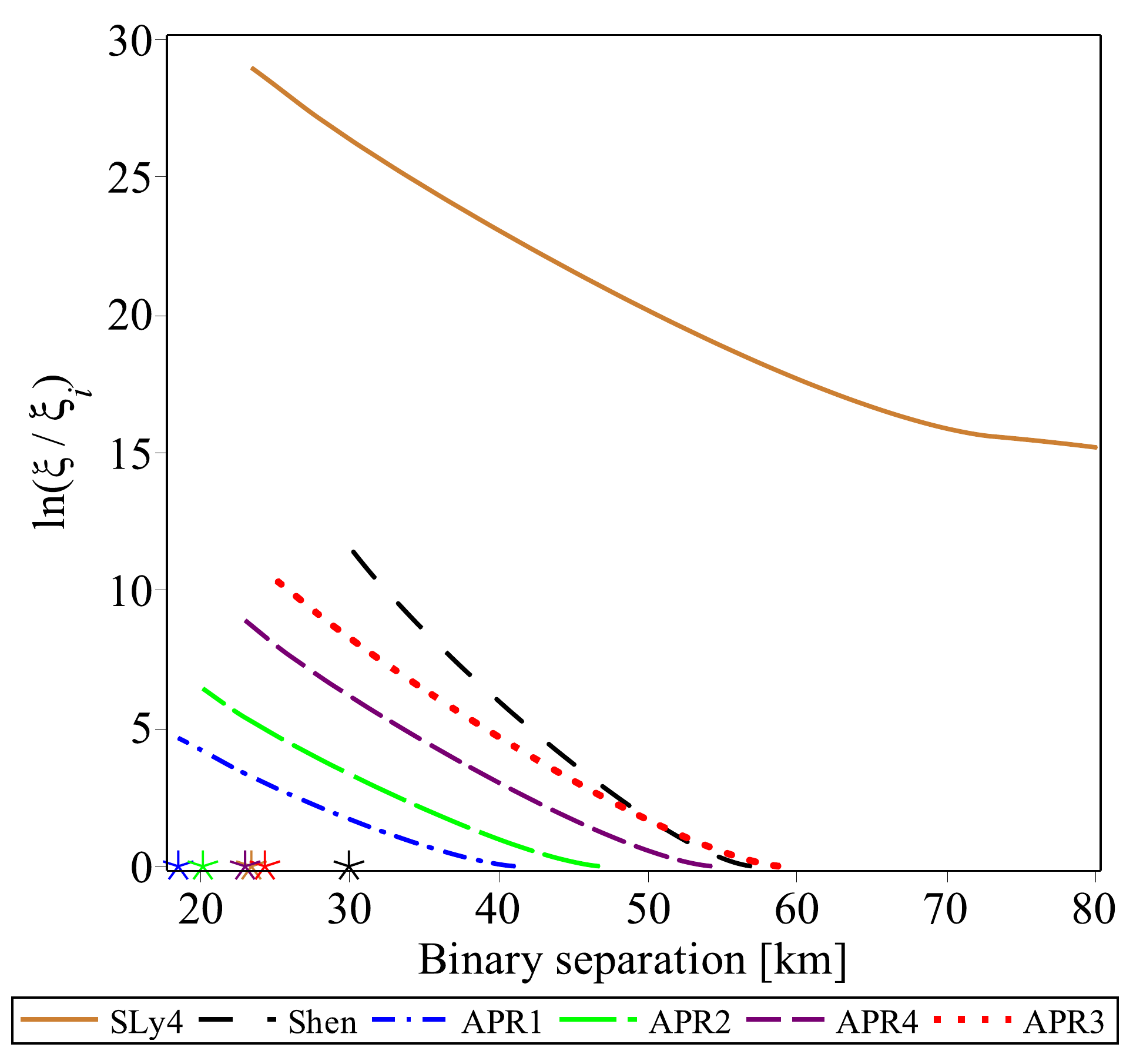}
\end{overpic}
\caption{The relationship between the binary separation and the energy $e$-folding number, with $\xi$ denoting the modal oscillation amplitude and $\xi_i$ its initial value at the instability threshold. Positions marked by asterisks indicate the distance at which the binary merges. Our results should be compared with similar energy $e$-folding diagrams presented in W2016, see for example their Fig.~13 for results assuming very high order $(n\geq 1000)$ g-modes within the SLy4 EOS.}
\label{fig:growth}
\end{figure}

  For concreteness, the discussion in this section specializes to a specific $l_g=4$ example g-mode, whose frequency, together with the MTCS values, are tabulated in Tb.~\ref{tb:nstMTCS} for six EOS choices. We see from Tb.~\ref{tb:nstMTCS} that with most EOSs, the ${\rm MTCS}$ value is roughly on the order of $10^{-2}$, as compared to the typical size of $\sim 10^{-5}$ (Tb.~\ref{tb:staticMTCS}, and also Tb.~\ref{tb:nstMTCS}) under a static tide. This makes the onset of instabilities possible in a binary coalescence scenario. Namely, some perturbed g-mode frequencies $\omega_-^2$ could become negative well before merger. To directly illustrate this effect, we estimate the instability threshold, the growth rate and the growth window for the example mode.
Aside from our MTCS results with this example mode however, we note in addition that in the case of very high order g-modes with even lower eigen-frequencies (e.g., $\omega_g \lesssim 1 \: \rm{rad/s}$), the three- and four-mode residual $R_{gg}$ can rise to overwhelmingly large values (W2016, Sect.~\ref{sec:nst_result}), which would further enhance instabilities, bringing its onset forward to an even earlier instant during inspiral.
  
  Instability begins when the square of the perturbed g-mode frequency crosses zero. The threshold separation when this occurs is listed in Tb.~\ref{tb:nstMTCS} for each EOS, which is estimated using linear interpolation between discrete MTCS data points. When the MTCS exceeds $1$, the perturbed frequency can be approximated by 
\begin{equation} \label{eq:growth}
\begin{aligned}
\omega_{-}\approx \pm i\omega_g\sqrt{{\rm MTCS}-1},
\end{aligned}
\end{equation}
and so the shifted modal frequency $\omega_-$ becomes imaginary, and exponentially drives the mode to large amplitudes with growth rate $\omega_g\sqrt{\rm MTCS-1}$ by appropriating energy from orbital motion. It is worth mentioning that the energy injection rate  $dE_{\rm inj} / dt$ from the tidal potential into the unstable g-mode is a crucial quantity, whose balance with factors such as the damping effects are vital for determining the instability window\footnote{See e.g.~Ref.~\cite{2015PhRvD..92h4018P} for derivations and illustrations of the instability window for the fundamental mode (f-mode, radial order $n=0$) in spinning neutron stars.}. Unfortunately, a lack of detailed knowledge regarding the driving process prevents an \emph{a priori} determination of $dE_{\rm inj}/dt$ within the scope of this paper.
 
  This diverging growth is not likely to be interrupted by the ``collapse instability'' (proposed by \cite{PhysRevLett.75.4161}, suggesting that coalescing neutron stars may collapse into black holes before merger) prior to merger because of the tidal stabilization effect \citep{1996PhRvL..76.4878L,1998PhRvD..58b3002S}. It would either be terminated by dissipative effects and so the modes saturate, or simply continues until merger happens (or possibly tear the neutron star apart before merger). 
A description of the saturation configuration is beyond the scope of this paper (refer to \cite{2016PhRvD..94j3012E} though for a first attempt on this topic), so we will begin by ignoring it and computing the maximum final modal amplitude potentially achievable before merger for each EOS. 
This quantity is limited by both the instability growth rate and the growth window. The $t_{\rm mg}$ in Tb.~\ref{tb:nstMTCS} denotes the approximate maximum growth window, defined as the temporal interval between the onset of instability and merger\footnote{For the estimation of $t_{\rm mg}$, we adopt \cite{1994MNRAS.270..611L} Eq.~2.12 which gives the orbital decay rate due to the emission of GWs from a point-mass binary. However, we caution that this formula suffers from complications during the late inspiral stage, partly from the so-called dynamical instability considered by \cite{1994ApJ...420..811L}. The dynamical instability indicates that circular orbits will become unstable when  $A\lesssim 3\mathcal{R}$ (the exact value depends on the EOS). With such close separations, tidal effects would accelerate the orbital decay rate, hence narrow the growth window of the unstable modes. Nevertheless, the discrepancy between the $t_{\rm mg}$ as given by \cite{1994MNRAS.270..611L} Eq.~2.12 and from more advanced formalisms is tolerable (see the middle panel of \cite{1994ApJ...420..811L} Fig.~4).}.

  These windows are quite narrow in general, but much wider for the SLy4 EOS (the inspiral duration depends on the initial ``onset separation" rather nonlinearly, a fact that can potentially magnify any observable signal's sensitivity on EOS differences). The exponential growth in modal amplitudes then further acts as a highly effective amplifier, with the final amplitude of unstable modes differing by more than {10} orders of magnitude between the SLy4 and the {APR1} EOS (see Fig.~\ref{fig:growth}). 

  Such massive differences can potentially lead to observable consequences, and a subsequent constraining of the EOS. For our discussion, we concentrate on the GWs given off by the binary system as opposed to the electromagnetic signature. This is because first of all, the GWs reveal the celestial mechanics of the neutron stars more cleanly, without contamination/blockage due to processes occurring within the magnetosphere or in the interstellar medium. Secondly, GW detectors are ``all-sky'' (have broad antenna patterns), and as such have a good chance of observing the late inspiral stage (without needing to be alerted by dramatic triggers such as the merger itself). Indeed, studies by e.g., \cite{1998MNRAS.299.1059A} have identified unstable modes as promising candidates to be studied with GW asteroseismology, as they may grow to such large amplitudes that they emit GWs detectable on Earth \citep{2011PrPNP..66..239A}. Moreover, even when the wave from the modes themselves are too weak, indirect modal effects imprinted onto the more prominent orbital-motion-generated waves should be observable. In the most optimistic scenario, unmissable qualitative distinctions such as neutron stars with SLy4 EOS being torn apart pre-merger while those EOSs with larger buoyancy frequencies remaining intact, will imprint EOS information onto the amplitude of the gravitational waveform. On the other hand, in the conservative scenario that the modes simply saturate, at similar final amplitudes for different EOSs, the modes would still impart phase modifications onto the orbital motion, and thus the cumulative phase of the gravitational waveform. As some EOSs would reach saturation significantly earlier than others, the \emph{accumulated} phase corrections may still provide useful information. More specifically, the accumulated phase error scales as the the $-3$rd power of the initial orbital frequency at which instability appears (WAB), and using Kepler's law for a very crude estimate, the total accumulated phase error scales with threshold separation as its $4.5$th order. Taking data from Tb.~\ref{tb:nstMTCS}, this translates into approximately $10^2$ times greater total phase error for the least stable EOS to the most stable one. Finally, we have only considered inviscid fluids in this study, while linear dampings may further enhance the instability (W2016), and if also EOS-dependent, make it easier to determine the actual EOS. 
  
\begin{table*}[thb]
\centering
\caption{{The MTCS evaluated with the $l_g=4$, $n=32$ example g-modes at $A=95$ km (no resonance occurs at this binary separation), together with the instability threshold and the duration of the instability growth window. } 
\label{tb:nstMTCS}}
{\begin{tabular*}{0.95\textwidth}{@{\extracolsep{\fill}}cccccc}
\toprule[2pt]
\multirow{2}{*}{EOS} & \multirow{2}{*}{$f_g$ (Hz)} & \multicolumn{2}{c}{MTCS ($A=95$ km)} & \multicolumn{2}{c}{Instability} \\

&  & Static tide & Non-static tide & Threshold (km) & $t_{\rm mg}$ (ms) \\
\midrule[1pt]
  SLy4 & $2.71$ & $1.75\times 10^{-3}$ & $2.01$ & $144$ & $1583$ \\ 
  
  Shen & $27.7$ & $1.60\times 10^{-4}$ & $1.23 \times 10^{-2}$ & $57.0$ & $35.4$ \\ 
 
  APR1 & $23.3$ & $4.53\times 10^{-5}$ & $9.71 \times 10^{-2}$ & $41.2$ & $10.1$ \\ 
 
  APR2 & $23.7$ & $2.99\times 10^{-5}$ & $5.27\times 10^{-2}$ & $46.8$ & $16.9$ \\ 
 
  APR3 & $15.9$ & $5.58\times 10^{-5}$ & $1.91\times 10^{-2}$ & $58.9$ & $42.8$ \\ 

  APR4 & $18.8$ & $3.95\times 10^{-5}$ & $4.10\times 10^{-2}$ & $54.4$ & $30.7$ \\
\bottomrule[2pt]
\end{tabular*}}
\end{table*}
  
\section{Conclusion} \label{sec:Conclusion}

  In this paper, we have examined the EOS-dependence of the growth of neutron star g-modes driven by tidal interactions. A representative collection of six EOSs at zero-temperature are considered. The twin neutron stars are assumed to consist of normal fluid, each with mass $M\approx 1.4\rm M_{\odot}$ and non-spinning. {The cases of both static and non-static tides are examined, using the volume-preserving transformation of VZH and the original formalism describing nonlinear mode interactions by WAQB, respectively.} With static tide, our results show that while the mode-tide coupling is stronger in stars that are larger in size and with smaller buoyancy frequencies, the near-exact cancellation between the three- and four- mode couplings as revealed by VZH provides a sufficient level of suppression that the g-modes remain stable with all the EOSs considered. In other words, the qualitative conclusion reached by VZH regarding the stability is explicitly shown to be EOS independent. On the other hand, {a return of the instability under non-static tides is unambiguously seen from our numerical results, even for moderately high-order p-g pairs.} 
 The combination of a longer growth window (the instability does not rely on any resonance condition being met, so once it appears, it continues to be present as the neutron stars inspiral) and a larger growth rate enables g-modes in stars with smaller buoyancy frequencies to either, (1) grow to much larger ({order of $\sim 10^{12}$ from its initial value, for the extreme case of SLy4}) 
amplitudes if mode saturation (or star disruption) does not occur or occurs only at very large amplitudes, (2) reach low amplitude saturation states much earlier. When these results for the pre-merger inspiral stage is viewed in conjunction with other studies mentioned in the introduction, and the strong EOS dependence seen for the post-merger remnant by \cite{2011MNRAS.418..427S}, there is reason for optimism that strategies for constraining the EOS using GW observations should be viable, utilizing features in the amplitude and phase of the waveforms to cater for the two scenarios above, respectively or in combination. 
  
  Nevertheless, there is still a long winding road that we need to traverse to go from our highly stylized derivations to realistic predictions for observational signatures. Many uncertainties that may affect the mode-tide coupling and the p-g instability remain untouched. First of all, neutron star's structural details, such as whether it is spinning or non-spinning, or if it has zero or finite temperature, should all influence the mode-tide coupling strength. Moreover, hydrodynamical effect would also influence the magnitude of the coupling constants. For instance, with the same EOS, but assuming either normal fluid or superfluid for the neutron star matter will result in different buoyancy frequencies (\cite{2017MNRAS.464.2622Y} Fig.~2), and thereby lead to different MTCS values (Sect.~\ref{sec:staticresult}). Even more urgently needed for the purpose of predicting the instability's impact on GW waveforms, is an understanding of the complications arising when the unstable modes grow to large amplitudes. As discussed in Sect.~\ref{sec:instability}, this includes, but is not limited to: the determination of the instability window, a quantitative analysis of (both the linear and the nonlinear) damping, and the saturation of the unstable g-modes.

  On the other hand, the MTCS's sensitivity to the buoyancy frequency also reveals the limitation of the tabulated $\rho -P$ EOS data that are currently available in literature, and the demand for a more thorough description of nuclear matter with ultra-high densities, including its composition and properties away from $\beta$-equilibrium.
Further analytical investigations would likely be very involved, as more complications arise when modes grow to large amplitudes. For example, the four-mode interactions included in the computations here and the previous literature include only two daughter eigen-modes interacting with two copies of the tidal field, but interactions involving three or more eigen-modes would have to be considered if these modes grow to large amplitudes. An important step in terms of further characterization of the instability issue would thus possibly be the development of numerical simulations beginning sufficiently early in the inspiral for there to be enough time for the modes to grow (WAB, W2016), which explore a variety of EOSs (see e.g.~\cite{2017AdAst2017E...1S} for numerical efforts in this direction). 
If more precise data are needed in the future though, more sophisticated analytical treatments should be possible on this front.

\acknowledgments
We thank Nevin Weinberg for kindly and patiently answering many questions regarding the computation of the linear tide and the usage of the \textsc{adipls} package, his valuable comments on this manuscript is also greatly appreciated. FZ is supported by NSFC Grants 11443008 and 11503003, Fundamental Research Funds for the Central Universities Grant No.~2015KJJCB06, and a Returned Overseas Chinese Scholars Foundation grant.

  \appendix
  \renewcommand{\appendixname}{Appendix~\Alph{section}}

\section{A. The Jacobians} \label{app:Jaco}
  
  The full expression for the perturbed g-mode frequency in the initial coordinate system is given by Eq.~\eqref{eq:perturbedgmode1} (VZH Eq.~B3). After the VPT, the equivalent expression consists of Jacobians and the MTCS, and is given by Eq.~\eqref{eq:perturbedgmode2}. {Within static tides}, our attention is focused on the MTCS, but $\epsilon J_{gg}^{(1)}$ is nevertheless non-negligible in certain circumstances. Therefore, this appendix section is devoted to the calculation of $\epsilon J_{gg}^{(1)}$, and to give the reason why the last term of Eq.~\eqref{eq:perturbedgmode2} can be ignored.
  
  We begin with the expression for the first-order Jacobians as given in VZH Eq.~89,
\begin{equation} \label{eq:Jacobian}
J_{ab}^{(1)}=-\frac{\omega_a^2}{E_0}I_{ab\chi^{(1)}}\,,
\end{equation}
where the integral $I_{ab\chi^{(1)}}$ is given by VZH Eq.~81:
\begin{equation} \label{eq:integral}
I_{ab\chi^{(1)}} = \int dr \, r^2\rho \left[T a_r b_r\frac{d\chi_r^{(1)}}{dr}+F_a \frac{a_rb_h}{r}\left(\chi_r^{(1)} - \chi_h^{(1)}\right)+F_b b_r a_h\frac{d\chi_h^{(1)}}{dr}
+\frac{a_h b_h}{r}\left(G_{\chi^{(1)}}\chi_h^{(1)}+F_{\chi^{(1)}}\chi_r^{(1)}\right)\right],
\end{equation}
where $a_r$ and $a_h$ are the radial and horizontal components of the eigenfunction for mode \textit{a}, and
\begin{equation} \label{eq:chirh}
\begin{aligned}
\chi_r^{(1)}&=\frac{\omega_0^2r^2}{\mathfrak{g}},\\
\chi_h^{(1)}&=\frac{1}{r\, l(l+1)}\frac{d (r^2\chi_r^{(1)})}{dr},
\end{aligned}
\end{equation}
being the radial and horizontal components of the tidal deformation induced by a static tide, respectively. The angular integrals {$T$, $F_a$ and $G_a$ are defined as (WAQB Eq.~A20-A22, VZH Eq.~D1-D3)}
\begin{equation} \label{eq:angularint}
\begin{aligned}
T &= \int_0^{2\pi}\int_0^{\pi}Y_{l_a m_a}Y_{l_b m_b}Y_{lm}\sin\theta d\theta d\phi\,,\\
F_a&=\frac{T}{2}\left(\Lambda_b^2+\Lambda_{\chi^{(1)}}^2-\Lambda_a^2\right)\,,\\
G_a&=\frac{T}{4}\left[\Lambda_a^4-\left(\Lambda_b^2-\Lambda_{\chi^{(1)}}^2\right)^2\right]\,.
\end{aligned}
\end{equation}

  The expression for $\epsilon J_{gg}^{(1)}$ can be deduced from Eqs.~\eqref{eq:Jacobian} and \eqref{eq:integral}, which turns out to be
\begin{equation} \label{eq:Jgg}
\epsilon J_{gg}^{(1)} = -\frac{\mathcal{R}^3}{A^3}\frac{\omega_g^2}{E_0}\int_0^{\mathcal{R}} dr \, r^2\rho \left[T g_r^2\frac{d\chi_r^{(1)}}{dr}+F_g \frac{g_rg_h}{r}\left(\chi_r^{(1)}-\chi_h^{(1)}\right)+F_g g_r g_h\frac{d\chi_h^{(1)}}{dr}
+\frac{g_h^2}{r}\left(G_{\chi^{(1)}}\chi_h^{(1)}+F_{\chi^{(1)}}\chi_r^{(1)}\right)\right].
\end{equation}
After substituting in the values for $g_r$ and $g_h$ calculated in Sect.~\ref{sec:grgh}, and inserting the expressions for $\chi_r^{(1)}$ and $\chi_h^{(1)}$ from Eq.~\eqref{eq:chirh}, we are ready to evaluate Eq.~\eqref{eq:Jgg} and obtain an estimate of the size of the Jacobians. To be in accordance with the main text, we use the six EOSs introduced in Sect.~\ref{sec:6EOS} and set $l_g=4$, $l=2$. From Eq.~\eqref{eq:angularint}, we then obtain {$T=(10\sqrt{5})/(77\sqrt{\pi})$, $F_g=3T$, $F_{\chi^{(1)}}=17T$, $G_{\chi^{(1)}}=9T$} and ultimately arrive at the numerical results shown in Tb.~\ref{tb:staticMTCS}.

  Now we can also compare $\epsilon^2(J_{gg}^{(1)})^2$ with the MTCS. When $A=100 \: \rm km$, $\epsilon^2(J_{gg}^{(1)})^2\sim 10^{-8}$ with the six EOSs, at least $10^3$ times smaller than the MTCS (see Tb.~\ref{tb:staticMTCS}), thus it is reasonable to ignore the former. Moreover, since all first-order Jacobians are governed by Eq.~\eqref{eq:Jacobian}, one can confirm that $2\epsilon^2J_{pg}^{(1)}J_{gp}^{(1)}$ and $\epsilon^2(J_{gp}^{(1)})^2$ are also much smaller than the MTCS, by repeating the same computational procedures for other modes. In addition, as stated in VZH Sect.~4.2, the second order Jacobians are of the same magnitude as the first-order ones. Therefore, $2\epsilon^2J_{gg}^{(2)}$ is also a small quantity. All in all, it is safe to ignore all Jacobian terms at $\mathcal{O}(\epsilon^2)$.

\section{B. Review of the Volume Preserving Transformation} \label{app:VPT}

  The VPT developed by Venumadhav, Zimmerman and Hirata (VZH) plays {an important} role in our analysis. Therefore, in this appendix, we review it in the context of static tide first, and then partially extend the transformation to non-static tide. Our introduction is only cursory and a thorough discussion can be found in VZH Sect.~3.1.
  
  In the initial coordinate system $(r,\theta,\phi)$, the star is deformed by the tidal force to an irregular shape, hence the tidal displacement depends on $r$, $\theta$ and $\phi$. Fortunately, there (always) exists another coordinate system $(R,\Theta,\Phi)$, in which the deformed star regains its spherical symmetry. This coordinate system, once found, will greatly simplify the computation of the mode-tide coupling. This is the aim of the VPT. Representing this transformation by an infinitesimal displacement vector $\vec{\zeta}$, the physical constraint it must satisfy is that the volume must be conserved, or
\begin{equation} \label{eq:div0}
\begin{aligned}
\nabla\cdot\vec{\zeta}=0.
\end{aligned}
\end{equation}
That $\vec{\zeta}$ is divergence-free and that the physical quantities should not be affected by the choice of coordinate systems are the two crucial conditions for the VPT program. The first one is the only constraint on displacement vector $\vec{\zeta}$ while the second can be applied to the (tidally perturbed) gravitational potential, whose expression, in the static tide scenario, is
\begin{equation} \label{eq:gravpot}
\begin{aligned}
\Psi&=\Psi_0(r)+\epsilon U_{\rm st}=\Psi_0(r)-\epsilon\omega_0^2r^2W_{20}Y_{20}(\theta),
\end{aligned}
\end{equation}
where $\Psi_0(r)$ is the potential of the unperturbed star, and $U_{\rm st}$ is the static tidal potential ($m=0$ term in Eq.~\eqref{eq:dyntide}). Eq.~\eqref{eq:gravpot} implicitly suggests that the Cowling approximation is employed, i.e., the perturbation from the star itself is ignored in the expression of $\Psi$. The VPT is effective by the infinitesimal displacement operator $\mathcal{D}$, via
\begin{equation} \label{eq:VPTdisplacement}
\begin{aligned}
\vec{X}&=\mathcal{D}(\vec{\zeta}(\vec{x}))\ket{\vec{x}}=\ket{\vec{x}+\vec{\zeta}(\vec{x})}\,,\\
\vec{x}&=\mathcal{D}(-\vec{\zeta}(\vec{X}))\ket{\vec{X}}=\ket{\vec{X}-\vec{\zeta}(\vec{X})}\,,
\end{aligned}
\end{equation}
where $\vec{x}$ and $\vec{X}$ stand for the coordinate systems before and after the VPT respectively. In Eq.~\eqref{eq:VPTdisplacement} we have followed the notations from quantum mechanics, where $\mathcal{D}(d\vec{x})$ simply represents the infinitesimal displacement operator that brings state $\ket{\vec{x}}$ to state $\ket{\vec{x}+d\vec{x}}$. Applying Eq.~\eqref{eq:VPTdisplacement} to $(R,\Theta)$, at leading order we have that
\begin{equation} \label{eq:RThetadisplacement}
\begin{aligned}
r&=R-\epsilon\vec{\zeta}^{(1)}\cdot\hat{R},\\
\theta&=\Theta-\epsilon\vec{\zeta}^{(1)}\cdot\hat{\Theta},
\end{aligned}
\end{equation}
where $\hat{R}$, $\hat{\Theta}$ are the unit basis vectors in the $R$ and $\Theta$ directions, and $\vec{\zeta}^{(1)}$ is the first order infinitesimal displacement (we limit our discussion to the leading order, while the derivation of $\vec{\zeta}^{(2)}$ is available in VZH Sect.~3.1). Recalling that the choice of coordinate systems would not affect the value of the scalar field $\Psi$, we can substitute Eq.~\eqref{eq:RThetadisplacement} into Eq.~\eqref{eq:gravpot}, obtaining
\begin{equation} \label{eq:VPTgrav1}
\begin{aligned}
\Psi=\Psi_0(R)-\epsilon\left[\mathfrak{g}\vec{\zeta}^{(1)}\cdot\hat{R}+\omega_0^2R^2W_{20}Y_{20}(\Theta)\right]+\mathcal{O}(\epsilon^2).
\end{aligned}
\end{equation}
Eq.~\eqref{eq:VPTgrav1} is similar to VZH Eq.~38, with the gravitational acceleration given by $\mathfrak{g}\equiv d\Phi_0/dR$, while $\vec{\zeta}^{(1)}\cdot\hat{R}$ is derived under the divergence-free constraint on $\vec{\zeta}$ that we mentioned earlier. The specific steps towards $\vec{\zeta}^{(1)}\cdot\hat{R}$ are expounded below.

  We begin with the spherical harmonic expansion of an arbitrary vector $\vec{E}$:
\begin{equation} \label{eq:sharmonic}
\begin{aligned}
\vec{E}=\sum_{l=0}^{\infty}\sum_{m=-l}^{l}\left(E_{1lm}(r)\vec{Y}_{lm}+E_{2lm}(r)\vec{\Psi}_{lm}+E_{3lm}(r)\vec{\Phi}_{lm}\right),
\end{aligned}
\end{equation}
with $\vec{Y}_{lm}=Y_{lm}\hat{r}$, $\vec{\Psi}_{lm}=r\nabla Y_{lm}$, and $\vec{\Phi}=\vec{r}\times\nabla Y_{lm}$. The divergence of this vector is
\begin{equation} \label{eq:divE}
\begin{aligned}
\nabla\cdot\vec{E}=\sum_{l=0}^{\infty}\sum_{m=-l}^{l}\left(\frac{dE_{1lm}}{dr}+\frac{2}{r}E_{1lm}-\frac{l(l+1)}{r}E_{2lm}\right)Y_{lm}.
\end{aligned}
\end{equation}
Combining Eqs.~\eqref{eq:div0} and \eqref{eq:divE},  $\forall \,l$ and $m=-l,\cdots,l$, we have that 
\begin{equation} \label{eq:divfree}
\begin{aligned}
\frac{d\zeta_{1lm}(R)}{dR}+\frac{2}{R}\zeta_{1lm}(R)-\frac{l(l+1)}{R}\zeta_{2lm}(R)=0.
\end{aligned}
\end{equation}
Once we define 
\begin{equation}
\begin{aligned}
\zeta_{2lm}\equiv \frac{u_{lm}}{R}+\partial_Ru_{lm},
\end{aligned}
\end{equation}
where $u_{lm}$ is a coefficient that depends only on $R$, the divergence-free condition \eqref{eq:divfree} then implies
\begin{equation}
\begin{aligned}
\zeta_{100}&=CR^{-2}\quad(l=0),\\
\zeta_{1lm}&=\frac{l(l+1)}{R}u_{lm}\quad(l\neq 0).
\end{aligned}
\end{equation}
Here we note that the static tidal potential is axi-symmetric, i.e. independent of the azimuth angle. Therefore, the displacement vector $\vec{\zeta}$ should also respect rotational symmetry. Substituting $u_{lm}$ into the spherical harmonic expansion of $\vec{\zeta}$ (Eq.~\eqref{eq:sharmonic}) and set $m=0$, we get
\begin{equation} \label{eq:uandzeta}
\begin{aligned}
\vec{\zeta}&=\sum_{l=0}^{\infty}\left(\zeta_{1l0}(R)\vec{Y}_{l0}+\zeta_{2l0}(R)R\nabla Y_{l0}\right)\\
&=CR^{-2}Y_{00}\hat{R}+\sum_{l=1}^{\infty}\left[\frac{l(l+1)}{R}u_{l0}Y_{l0}(\Theta)\hat{R}+\left(\frac{u_{l0}}{R}+\partial_Ru_{l0}\right)\partial_{\Theta}(Y_{l0}(\Theta))\hat{\Theta}\right],
\end{aligned}
\end{equation}
which is always divergence-free. Now that we have the radial component of $\vec{\zeta}$ from Eq.~\eqref{eq:uandzeta}, Eq.~\eqref{eq:VPTgrav1} then turns out to be
\begin{equation} \label{eq:VPTgrav2}
\begin{aligned}
\Psi=\Psi_0(R)-\epsilon\left[\mathfrak{g}\left(CR^{-2}Y_{00}+\sum_{l=1}^{\infty}\frac{l(l+1)}{R}u_{l0}Y_{l0}(\Theta)\right)+\omega_0^2R^2W_{20}Y_{20}(\Theta)\right]+\mathcal{O}(\epsilon^2)\,.
\end{aligned}
\end{equation}
The gravitational potential being finite at $R=0$ requires that $C=0$. We (as VZH did) then insist that $\Psi$ in the new coordinate system depends only on $R$ (all terms that contain the spherical harmonic function must be eliminated), yielding (VZH Eq.~45)
\begin{equation}
\begin{aligned}
u_{l0}^{(1)}=-\frac{\omega_0^2R^3W_{20}}{6\mathfrak{g}}\delta_{l2}\,.
\end{aligned}
\end{equation}
Finally, placing $u_{l0}^{(1)}$ back into Eq.~\eqref{eq:VPTgrav2}, we immediately confirm that the first order external potential (also the first order radial displacement) vanishes.

  Now we move on to non-static tide. The only difference between static and non-static tides lies in the form of the tidal potential, with the latter characterized by an exponential factor $e^{-im\Omega t}$, hence depends on both space and time. This time-dependence in the potential term leads to a profound physical effect, and will influence the VPT at a fundamental level. To be more specific, when conducting the transformation, one should not only consider an infinitesimal (and volume-preserving) displacement of fluid elements, but also take the infinitesimal time evolution into account. Otherwise, if time is frozen, the exponential factor shall reduce to a constant and there will be no difference between the static and non-static tidal potentials. Actually, for inspiraling binary systems with quasi-circular orbits, the effect of infinitesimal time evolution on the subject star is equivalent to a rotation of an infinitesimal angle $\Omega\Delta t$ for a small fluid parcel at position $\vec{x}(r,\theta,\phi)$ on that star ($\Delta t$ symbolizes the finite time interval). In other words, the whole transformation is now
\begin{equation}
\begin{aligned}
(r,\theta,\phi)\xrightarrow[\rm displacement]{\rm{infinitesimal}}(r^{\prime},\theta^{\prime},\phi^{\prime})\xrightarrow[\rm (rotation)]{\rm time\:evolution} (R,\Theta,\Phi).
\end{aligned}
\end{equation}
In the form of operators and the bra-ket notation (analogous to Eq.~\eqref{eq:VPTdisplacement}),
\begin{equation}  \label{eq:nstVPTdis}
\begin{aligned}
\vec{X}&=\mathcal{H}(\Delta t)\mathcal{D}(\vec{\zeta}(\vec{x}))\ket{\vec{x}}=\mathcal{R}(\Omega\Delta t)\mathcal{D}(\vec{\zeta}(\vec{x}))\ket{\vec{x}}=\ket{\vec{x}+\vec{\zeta}(\vec{x})+\vec{\delta}(\vec{x})}\,,\\
\vec{x}&=\mathcal{D}(-\vec{\zeta}(\vec{X}))\mathcal{R}(-\Omega\Delta t)\ket{\vec{X}}=\ket{\vec{X}-\vec{\zeta}(\vec{X})-\vec{\delta}(\vec{X})}\,,
\end{aligned}
\end{equation}
where $\vec{\delta}$ denotes the displacement induced by rotation. Relation \eqref{eq:nstVPTdis} represents the displacement operator $\mathcal{D}$ and the temporal evolution operator $\mathcal{H}$ (equivalently, the rotation operator $\mathcal{R}$\footnote{The origin of our chosen coordinate system is the center of the subject star, with the orbital plane being the $x-y$ plane. Thus, the precise effect of $\mathcal{R}(\Omega\Delta t)$ is to rotate states around the $z$-axis by an angle $\Omega\Delta t$.}) acting on a small fluid parcel with state $\ket{\vec{x}}$. Apply the transformation above to $R$ (we do not explicitly write out similar relations for the $\Theta$- and $\Phi$-components since they are not needed in the ensuing derivations), we obtain to leading order that 
\begin{equation}
\begin{aligned}
r&=R-\epsilon\vec{\zeta}^{(1)}\cdot\hat{R}+\frac{1}{2}R\sin^2(\Theta)(\Omega\Delta t)^2.\\
\end{aligned}
\end{equation}
The next step is to express the gravitational potential in the new coordinate system. Since the procedure is nearly identical to the static case, we skip the derivation details and write down $\Psi(R,\Theta,\Phi,t)$ directly:
\begin{equation} \label{eq:nstgrav}
\begin{aligned}
\Psi(R,\Theta,\Phi,t)=&\Psi_0(R)-\epsilon\mathfrak{g}\vec{\zeta}^{(1)}\cdot\hat{R}+\frac{1}{2}\mathfrak{g}R\sin^2(\Theta)(\Omega\Delta t)^2-\epsilon\omega_0^2R^2W_{22}Y_{22}(\Theta,\Phi)e^{-i2\Omega t} \\
=&\Psi_0(R)-\epsilon\mathfrak{g}\vec{\zeta}^{(1)}\cdot\hat{R}+\frac{1}{3}\mathfrak{g}R(\Omega\Delta t)^2-\frac{2}{3}\sqrt{\frac{\pi}{5}}\mathfrak{g}RY_{20}(\Theta)(\Omega\Delta t)^2-\epsilon\omega_0^2R^2W_{22}Y_{22}(\Theta,\Phi)e^{-i2\Omega t}.
\end{aligned}
\end{equation}
Comparing Eq.~\eqref{eq:VPTgrav1} with \eqref{eq:nstgrav}, it is manifest that new physics within the non-static tide context lie in the orbital angular frequency $\Omega$. This observation echoes the one made through Eq.~\eqref{eq:divchi}, but is made from the perspective of the VPT. In order to eliminate the angular dependence of $\Psi$ in the new coordinate system, the expression of $\vec{\zeta}^{(1)}\cdot\hat{R}$ should be
\begin{equation}
\begin{aligned}
\vec{\zeta}^{(1)}\cdot\hat{R}=\frac{2}{3}\sqrt{\frac{\pi}{5}}\frac{(\Omega\Delta t)^2}{\epsilon}RY_{20}(\Theta)-\frac{\omega^2R^2W_{22}}{\mathfrak{g}}Y_{22}(\Theta,\phi)e^{-i2\Omega t}.
\end{aligned}
\end{equation}
Plugging the expression above into Eq.~\eqref{eq:nstgrav}, we finally obtain
\begin{equation}
\begin{aligned}
\Psi=\Psi_0(R)+\frac{1}{3}\mathfrak{g}R(\Omega\Delta t)^2,
\end{aligned}
\end{equation}
with $\mathfrak{g}R(\Omega\Delta t)^2/3$ being the external potential. It is clear that due to the appearance of a nonvanishing orbital frequency, the non-static tidal potential dose not preserve the volume of a star even in leading order. 

  Up to now, we have seen the VPT with both static and non-static tidal potentials, and have explained the emergence of the finite orbital frequency correction to the gravitational potential (the main assertion in W2016) at an illustrative level. However, note that both $\vec{\zeta}^{(1)}\cdot\hat{R}$ and the external potential depend on the value of the finite time interval $\Delta t$, which can be of any arbitrary value. Hence, neither the external potential nor the expression for $\vec{\zeta}^{(1)}\cdot\hat{R}$ derived above can be utilized in our numerical evaluations. {To this end, we adopt the direct method developed by WAQB in the main text of the paper, that allows us to compute the MTCS explicitly.}

\section{C. Numerical method for the non-static linear tide} \label{app:numchi}

  The linear tidal displacement $\vec{\chi}^{(1)}$ represents the response of a fluid star to the external non-static tidal field, which, as shown in Sect.~\ref{sec:roadtoMTCS}, couples nonlinearly to the normal modes of neutron star and result in the frequency shift. It is outlined in Sect.~\ref{sec:fsft_new} that the (non-static) linear tide is solved numerically by means of the so-called shooting technique. This appendix, as a supplementary section to Sect.~\ref{sec:fsft_new}, serves for a detailed description of our numerical approach.
  
  $\vec{\chi}^{(1)}$ is governed by the forced oscillation equations with the $m=\pm 2$ tidal potential being the driving term. The standard form of this ODE system can be expressed as (\cite{2008ApJ...679..783P} Eqs.~(A1)-(A4), WAQB Eqs.~(A9)-(A11))
\begin{equation} \label{eq:forcedODE}
\begin{aligned}
\frac{1}{r^2}\frac{d}{dr}(r^2\chi_r^{(1)})-\frac{\mathfrak{g}}{c_s^2}\chi_r^{(1)}+\left(1-\frac{l(l+1)c_s^2}{r^2(m\Omega)^2}\right)\frac{P^{\prime}}{\rho c_s^2}-\frac{l(l+1)}{r^2(m\Omega)^2}\Phi^{\prime}&=-\epsilon\frac{l(l+1)\omega_0^2W_{22}}{(m\Omega)^2}, \\
\frac{1}{\rho}\frac{dP^{\prime}}{dr}+\frac{\mathfrak{g}}{\rho c_s^2}P^{\prime}+\left(N^2-(m\Omega)^2\right)\chi_r^{(1)}+\frac{d\Phi^{\prime}}{dr}&=2\epsilon\omega_0^2rW_{22}, \\
\frac{1}{r^2}\frac{d}{dr}\left(r^2\frac{d\Phi^{\prime}}{dr}\right)-\frac{l(l+1)}{r^2}\Phi^{\prime}-4\pi G\rho \left(\frac{P^{\prime}}{\rho c_s^2}+\frac{N^2}{\mathfrak{g}}\chi_r^{(1)}\right)&=0,
\end{aligned}
\end{equation}
where $\Phi^{\prime}$ is the Eulerian perturbation to the gravitational potential of the star itself, $\Omega$ being the orbital angular frequency and $l=2$, $m=\pm 2$ for the non-static tide. The inner and surface boundary conditions for \eqref{eq:forcedODE} are the same as for free oscillation (i.e., conditions \eqref{eq:inner_boundary} and \eqref{eq:surface_boundary}).

  Now we have presented the boundary value problem, the subsequent step is to reduce the standard oscillation equations \eqref{eq:forcedODE} to a dimensionless formulation, which is more suitable for numerical computation. Following the conventions in \cite{2008Ap&SS.316..113C}, we set
\begin{equation} \label{eq:dimlessy}
\begin{aligned}
y_1(x)&=\frac{\chi_r^{(1)}}{\mathcal{R}}, \\
y_2(x)&=x\left(\frac{P^{\prime}}{\rho}+\Phi^{\prime}\right)\frac{l(l+1)}{(m\Omega)^2r^2},\\
y_3(x)&=-x\frac{\Phi^{\prime}}{\mathfrak{g}r}, \\
y_4(x)&=x^2\frac{d}{dx}\left( \frac{y_3}{x} \right),
\end{aligned}
\end{equation}
in which we have introduced the (dimensionless) radius fraction $x\equiv r/\mathcal{R}$. Substituting the dimensionless quantities $y_1\ldots y_4$ into \eqref{eq:forcedODE}, the original fourth order ODE turns into
\begin{equation} \label{eq:dimlessODE}
\begin{aligned}
&x\frac{dy_1}{dx}+\left(2-\frac{\mathfrak{g}r}{c_s^2}\right)y_1+\left[\frac{(m\Omega)^2r^2}{l(l+1)c_s^2}-1\right]y_2+\frac{\mathfrak{g}r}{c_s^2}y_3=-\epsilon\frac{l(l+1)W_{22}x}{\varpi^2}, \\
&x\frac{dy_2}{dx}+l(l+1)\left[\frac{N^2}{(m\Omega)^2}-1\right]y_1+\left(1-\frac{N^2r}{\mathfrak{g}}\right)y_2-\frac{l(l+1)N^2}{(m\Omega)^2}y_3=\epsilon\frac{2l(l+1)W_{22}x}{\varpi^2}, \\
&x\frac{dy_3}{dx}-y_3-y_4=0, \\
&x\frac{dy_4}{dx}+\frac{4\pi G\rho r^2N^2}{\mathfrak{g}^2}y_1+\frac{4\pi G\rho r^3(m\Omega)^2}{l(l+1)\mathfrak{g}c_s^2}y_2-\left[l(l+1)+\frac{4\pi G\rho r}{\mathfrak{g}}\left(\frac{N^2r}{\mathfrak{g}}-2+\frac{\mathfrak{g}r}{c_s^2}\right)\right]y_3 +2\left(\frac{4\pi G\rho r}{\mathfrak{g}}-1\right)y_4=0,
\end{aligned}
\end{equation}
accompanied by the boundary conditions
\begin{equation} \label{eq:dimlessbound}
\begin{aligned}
y_2(0)&=(l+1)y_1(0), \\
y_4(0)&=0, \\
y_2(1)&=\frac{l(l+1)}{\varpi^2}\left(y_1(1)-y_3(1)\right), \\
y_4(1)&=-ly_3(1),
\end{aligned}
\end{equation}
where $\varpi$, defined via $\varpi\equiv m\Omega/\omega_0$, is called the dimensionless frequency ratio.

  Eqs.~\eqref{eq:dimlessODE} and \eqref{eq:dimlessbound} comprise a (purely mathematical) two point boundary value problem which is solved by the shooting method, for it is considerably more stable than other algorithms if one (or both) boundary is a singular or near-singular point. The essential concept of the shooting technique (a comprehensive introduction to the shooting method is available in \cite{1992nrfa.book.....P}) is to convert the boundary value problem into two initial value problems. One then specifies all initial conditions and then start to integrate (shoot) Eq.~\eqref{eq:dimlessODE} from the center and the surface, respectively, to an intermediate fitting point. The mismatch between the two asymptotes at the fitting point can be gradually reduced with Newton-Raphson iterations until a tolerable error is reached. The thus-calculated radial component of the linear tide (scaled) is displayed in Fig.~\ref{fig:chi_r}.

\bibliographystyle{apj}
\bibliography{References} 

\begin{thebibliography}{}
\expandafter\ifx\csname natexlab\endcsname\relax\def\natexlab#1{#1}\fi

\bibitem[{{Abbott} {et~al.}(2016{\natexlab{a}}){Abbott}, {Abbott}, {Abbott},
  {Abernathy}, {Acernese}, {Ackley}, {Adams}, {Adams}, {Addesso}, {Adhikari},
  \& et~al.}]{2016PhRvL.116x1103A}
{Abbott}, B.~P., {Abbott}, R., {Abbott}, T.~D., {et~al.} 2016{\natexlab{a}},
  Physical Review Letters, 116, 241103

\bibitem[{{Abbott} {et~al.}(2016{\natexlab{b}}){Abbott}, {Abbott}, {Abbott},
  {Abernathy}, {Acernese}, {Ackley}, {Adams}, {Adams}, {Addesso}, {Adhikari},
  \& et~al.}]{2016PhRvL.116f1102A}
---. 2016{\natexlab{b}}, Physical Review Letters, 116, 061102

\bibitem[{{Abbott} {et~al.}(2016{\natexlab{c}}){Abbott}, {Abbott}, {Abbott},
  {Abernathy}, {Acernese}, {Ackley}, {Adams}, {Adams}, {Addesso}, {Adhikari},
  \& et~al.}]{2016ApJ...832L..21A}
---. 2016{\natexlab{c}}, \apjl, 832, L21

\bibitem[{{Akmal} {et~al.}(1998){Akmal}, {Pandharipande}, \&
  {Ravenhall}}]{1998PhRvC..58.1804A}
{Akmal}, A., {Pandharipande}, V.~R., \& {Ravenhall}, D.~G. 1998, \prc, 58, 1804

\bibitem[{{Andersson}(2011)}]{2011PrPNP..66..239A}
{Andersson}, N. 2011, Progress in Particle and Nuclear Physics, 66, 239

\bibitem[{{Andersson} \& {Kokkotas}(1998)}]{1998MNRAS.299.1059A}
{Andersson}, N., \& {Kokkotas}, K.~D. 1998, \mnras, 299, 1059

\bibitem[{{Baym} {et~al.}(1971{\natexlab{a}}){Baym}, {Bethe}, \&
  {Pethick}}]{1971NuPhA.175..225B}
{Baym}, G., {Bethe}, H.~A., \& {Pethick}, C.~J. 1971{\natexlab{a}}, Nuclear
  Physics A, 175, 225

\bibitem[{{Baym} {et~al.}(1971{\natexlab{b}}){Baym}, {Pethick}, \&
  {Sutherland}}]{1971ApJ...170..299B}
{Baym}, G., {Pethick}, C., \& {Sutherland}, P. 1971{\natexlab{b}}, \apj, 170,
  299

\bibitem[{{Canuto}(1974)}]{1974ARA&A..12..167C}
{Canuto}, V. 1974, \araa, 12, 167

\bibitem[{{Chabanat} {et~al.}(1997){Chabanat}, {Bonche}, {Haensel}, {Meyer}, \&
  {Schaeffer}}]{1997NuPhA.627..710C}
{Chabanat}, E., {Bonche}, P., {Haensel}, P., {Meyer}, J., \& {Schaeffer}, R.
  1997, Nuclear Physics A, 627, 710

\bibitem[{{Chabanat} {et~al.}(1998){Chabanat}, {Bonche}, {Haensel}, {Meyer}, \&
  {Schaeffer}}]{1998NuPhA.635..231C}
---. 1998, Nuclear Physics A, 635, 231

\bibitem[{{Christensen-Dalsgaard}(2008)}]{2008Ap&SS.316..113C}
{Christensen-Dalsgaard}, J. 2008, \apss, 316, 113

\bibitem[{Christensen-Dalsgaard(2014)}]{CDBook}
Christensen-Dalsgaard, J. 2014, Lecture Notes on Stellar Oscillations
  (unpublished)

\bibitem[{{Damour} {et~al.}(2012){Damour}, {Nagar}, \&
  {Villain}}]{2012PhRvD..85l3007D}
{Damour}, T., {Nagar}, A., \& {Villain}, L. 2012, \prd, 85, 123007

\bibitem[{{Demorest} {et~al.}(2010){Demorest}, {Pennucci}, {Ransom}, {Roberts},
  \& {Hessels}}]{2010Natur.467.1081D}
{Demorest}, P.~B., {Pennucci}, T., {Ransom}, S.~M., {Roberts}, M.~S.~E., \&
  {Hessels}, J.~W.~T. 2010, \nat, 467, 1081

\bibitem[{{Duez} {et~al.}(2010){Duez}, {Foucart}, {Kidder}, {Ott}, \&
  {Teukolsky}}]{2010CQGra..27k4106D}
{Duez}, M.~D., {Foucart}, F., {Kidder}, L.~E., {Ott}, C.~D., \& {Teukolsky},
  S.~A. 2010, Classical and Quantum Gravity, 27, 114106

\bibitem[{{Essick} {et~al.}(2016){Essick}, {Vitale}, \&
  {Weinberg}}]{2016PhRvD..94j3012E}
{Essick}, R., {Vitale}, S., \& {Weinberg}, N.~N. 2016, \prd, 94, 103012

\bibitem[{{Flanagan} \& {Hinderer}(2008)}]{2008PhRvD..77b1502F}
{Flanagan}, {\'E}.~{\'E}., \& {Hinderer}, T. 2008, \prd, 77, 021502

\bibitem[{{Fuller} \& {Lai}(2011)}]{2011MNRAS.412.1331F}
{Fuller}, J., \& {Lai}, D. 2011, \mnras, 412, 1331

\bibitem[{{Gambhir} {et~al.}(1990){Gambhir}, {Ring}, \&
  {Thimet}}]{1990AnPhy.198..132G}
{Gambhir}, Y.~K., {Ring}, P., \& {Thimet}, A. 1990, Annals of Physics, 198, 132

\bibitem[{Harry(2010)}]{Harry:2010zz}
Harry, G.~M. 2010, Class.Quant.Grav., 27, 084006

\bibitem[{{Hebeler} {et~al.}(2013){Hebeler}, {Lattimer}, {Pethick}, \&
  {Schwenk}}]{2013ApJ...773...11H}
{Hebeler}, K., {Lattimer}, J.~M., {Pethick}, C.~J., \& {Schwenk}, A. 2013,
  \apj, 773, 11

\bibitem[{{Heiselberg} \& {Hjorth-Jensen}(2000)}]{2000PhR...328..237H}
{Heiselberg}, H., \& {Hjorth-Jensen}, M. 2000, \physrep, 328, 237

\bibitem[{{Hinderer} {et~al.}(2010){Hinderer}, {Lackey}, {Lang}, \&
  {Read}}]{2010PhRvD..81l3016H}
{Hinderer}, T., {Lackey}, B.~D., {Lang}, R.~N., \& {Read}, J.~S. 2010, \prd,
  81, 123016

\bibitem[{{Hotokezaka} {et~al.}(2013){Hotokezaka}, {Kyutoku}, \&
  {Shibata}}]{2013PhRvD..87d4001H}
{Hotokezaka}, K., {Kyutoku}, K., \& {Shibata}, M. 2013, \prd, 87, 044001

\bibitem[{{Lai}(1994)}]{1994MNRAS.270..611L}
{Lai}, D. 1994, \mnras, 270, 611

\bibitem[{{Lai}(1996)}]{1996PhRvL..76.4878L}
---. 1996, Physical Review Letters, 76, 4878

\bibitem[{{Lai} {et~al.}(1994){Lai}, {Rasio}, \&
  {Shapiro}}]{1994ApJ...420..811L}
{Lai}, D., {Rasio}, F.~A., \& {Shapiro}, S.~L. 1994, \apj, 420, 811

\bibitem[{{Lattimer}(2012)}]{2012ARNPS..62..485L}
{Lattimer}, J.~M. 2012, Annual Review of Nuclear and Particle Science, 62, 485

\bibitem[{{Lorenz} {et~al.}(1993){Lorenz}, {Ravenhall}, \&
  {Pethick}}]{1993PhRvL..70..379L}
{Lorenz}, C.~P., {Ravenhall}, D.~G., \& {Pethick}, C.~J. 1993, Physical Review
  Letters, 70, 379

\bibitem[{{Maselli} {et~al.}(2012){Maselli}, {Gualtieri}, {Pannarale}, \&
  {Ferrari}}]{2012PhRvD..86d4032M}
{Maselli}, A., {Gualtieri}, L., {Pannarale}, F., \& {Ferrari}, V. 2012, \prd,
  86, 044032

\bibitem[{{Pethick} {et~al.}(1995){Pethick}, {Ravenhall}, \&
  {Lorenz}}]{1995NuPhA.584..675P}
{Pethick}, C.~J., {Ravenhall}, D.~G., \& {Lorenz}, C.~P. 1995, Nuclear Physics
  A, 584, 675

\bibitem[{{Pfahl} {et~al.}(2008){Pfahl}, {Arras}, \&
  {Paxton}}]{2008ApJ...679..783P}
{Pfahl}, E., {Arras}, P., \& {Paxton}, B. 2008, \apj, 679, 783

\bibitem[{{Pnigouras} \& {Kokkotas}(2015)}]{2015PhRvD..92h4018P}
{Pnigouras}, P., \& {Kokkotas}, K.~D. 2015, \prd, 92, 084018

\bibitem[{{Press} {et~al.}(1992){Press}, {Teukolsky}, {Vetterling}, \&
  {Flannery}}]{1992nrfa.book.....P}
{Press}, W.~H., {Teukolsky}, S.~A., {Vetterling}, W.~T., \& {Flannery}, B.~P.
  1992, {Numerical recipes in FORTRAN. The art of scientific computing}

\bibitem[{{Read} {et~al.}(2013){Read}, {Baiotti}, {Creighton}, {Friedman},
  {Giacomazzo}, {Kyutoku}, {Markakis}, {Rezzolla}, {Shibata}, \&
  {Taniguchi}}]{2013PhRvD..88d4042R}
{Read}, J.~S., {Baiotti}, L., {Creighton}, J.~D.~E., {et~al.} 2013, \prd, 88,
  044042

\bibitem[{{Reisenegger} \& {Goldreich}(1992)}]{1992ApJ...395..240R}
{Reisenegger}, A., \& {Goldreich}, P. 1992, \apj, 395, 240

\bibitem[{{Shapiro} \& {Teukolsky}(1983)}]{1983bhwd.book.....S}
{Shapiro}, S.~L., \& {Teukolsky}, S.~A. 1983, {Black holes, white dwarfs, and
  neutron stars: The physics of compact objects}

\bibitem[{{Shen} {et~al.}(1998{\natexlab{a}}){Shen}, {Toki}, {Oyamatsu}, \&
  {Sumiyoshi}}]{1998NuPhA.637..435S}
{Shen}, H., {Toki}, H., {Oyamatsu}, K., \& {Sumiyoshi}, K. 1998{\natexlab{a}},
  Nuclear Physics A, 637, 435

\bibitem[{{Shen} {et~al.}(1998{\natexlab{b}}){Shen}, {Toki}, {Oyamatsu}, \&
  {Sumiyoshi}}]{1998PThPh.100.1013S}
---. 1998{\natexlab{b}}, Progress of Theoretical Physics, 100, 1013

\bibitem[{{Shibata} {et~al.}(1998){Shibata}, {Baumgarte}, \&
  {Shapiro}}]{1998PhRvD..58b3002S}
{Shibata}, M., {Baumgarte}, T.~W., \& {Shapiro}, S.~L. 1998, \prd, 58, 023002

\bibitem[{{Skyrme}(1959)}]{1959NucPh...9..615S}
{Skyrme}, T. 1959, \nphysa, 9, 615

\bibitem[{{Stergioulas} {et~al.}(2011){Stergioulas}, {Bauswein}, {Zagkouris},
  \& {Janka}}]{2011MNRAS.418..427S}
{Stergioulas}, N., {Bauswein}, A., {Zagkouris}, K., \& {Janka}, H.-T. 2011,
  \mnras, 418, 427

\bibitem[{{Suh} {et~al.}(2017){Suh}, {Mathews}, {Haywood}, \&
  {Lan}}]{2017AdAst2017E...1S}
{Suh}, I.-S., {Mathews}, G.~J., {Haywood}, J.~R., \& {Lan}, N.~Q. 2017,
  Advances in Astronomy, 2017, 612703

\bibitem[{{The Virgo Collaboration}(2012)}]{aVIRGO:2012}
{The Virgo Collaboration}. 2012, {Advanced Virgo Technical Design Report},
  {[VIR-0128A-12]}

\bibitem[{{Unno} {et~al.}(1989){Unno}, {Osaki}, {Ando}, {Saio}, \&
  {Shibahashi}}]{1989nos..book.....U}
{Unno}, W., {Osaki}, Y., {Ando}, H., {Saio}, H., \& {Shibahashi}, H. 1989,
  {Nonradial oscillations of stars}

\bibitem[{{Venumadhav} {et~al.}(2014){Venumadhav}, {Zimmerman}, \&
  {Hirata}}]{2014ApJ...781...23V}
{Venumadhav}, T., {Zimmerman}, A., \& {Hirata}, C.~M. 2014, \apj, 781, 23

\bibitem[{{Weinberg}(2016)}]{2016ApJ...819..109W}
{Weinberg}, N.~N. 2016, \apj, 819, 109

\bibitem[{{Weinberg} {et~al.}(2013){Weinberg}, {Arras}, \&
  {Burkart}}]{2013ApJ...769..121W}
{Weinberg}, N.~N., {Arras}, P., \& {Burkart}, J. 2013, \apj, 769, 121

\bibitem[{{Weinberg} {et~al.}(2012){Weinberg}, {Arras}, {Quataert}, \&
  {Burkart}}]{2012ApJ...751..136W}
{Weinberg}, N.~N., {Arras}, P., {Quataert}, E., \& {Burkart}, J. 2012, \apj,
  751, 136

\bibitem[{Wilson \& Mathews(1995)}]{PhysRevLett.75.4161}
Wilson, J.~R., \& Mathews, G.~J. 1995, Phys. Rev. Lett., 75, 4161

\bibitem[{{Witten}(1984)}]{1984PhRvD..30..272W}
{Witten}, E. 1984, \prd, 30, 272

\bibitem[{{Wu} \& {Goldreich}(2001)}]{2001ApJ...546..469W}
{Wu}, Y., \& {Goldreich}, P. 2001, \apj, 546, 469

\bibitem[{{Yu} \& {Weinberg}(2017)}]{2017MNRAS.464.2622Y}
{Yu}, H., \& {Weinberg}, N.~N. 2017, \mnras, 464, 2622

\end{thebibliography}

\end{document}